\documentstyle[epsf,psfig]{mn}
\def\mch{M$\rm^{c}$Hardy\,}
\def\ros{{\em ROSAT}\,}

\def\xmm{{\em XMM-Newton }}
\def\ch{{\em Chandra }}
\def\ecs{ergs cm$^{-2}$ s$^{-1}$~}
\def\ltsim{\mathrel{\hbox{\rlap{\hbox{\lower4pt\hbox{$\sim$}}}\hbox{$<$}}}}
\def\gtsim{\mathrel{\hbox{\rlap{\hbox{\lower4pt\hbox{$\sim$}}}\hbox{$>$}}}}
\def\etal{et al.~}
\def\etalc{et al.}
\def\ROSAT{{\em ROSAT }}
\def\Chandra{{\em Chandra }}
\def\XMM{{\em XMM }}
\def\aap{A\&A}
\def\aaps{A\&AS}

\def\apj{ApJ}
\def\apjl{ApJL}

\def\mnras{MNRAS}
\def\nat{Nat}

\begin{document}

\title[A {\em Chandra} Survey of the {\em XMM} 13hr Deep Field]
{A MEDIUM DEEP {\em CHANDRA} AND {\em SUBARU} SURVEY OF THE 13HR {\em XMM}/{\em
ROSAT} DEEP SURVEY AREA}

\author[M$\rm^{c}$Hardy, I.M., \etal]
{I. M. M$\rm^{c}$Hardy,$^{1}$ K. F.  Gunn,$^{1}$ 
A. M. Newsam,$^{2}$ K. O. Mason,$^{3}$ M. J. Page,$^{3}$
\newauthor T. Takata,$^{4}$ K. Sekiguchi,$^{4}$ T. Sasseen,$^{5}$ 
F. Cordova,$^{6}$ L. R. Jones$^{7}$ and N. Loaring$^{3}$\\
$^{1}$ Department of Physics and Astronomy, The University, Southampton SO17 1BJ\\
$^{2}$ Astrophysics Research Institute, Liverpool John Moores University,
Twelve Quays House, Egerton Wharf, Birkenhead CH41 1LD\\
$^{3}$ Mullard Space Science Laboratory, University College London, 
Holmbury St Mary, Dorking RH5 6NT\\
$^{4}$ National Astronomical Observatory of Japan, 650 North A`ohoku Place,
Hilo, HI 96720, USA\\
$^{5}$ Department of Physics, University of California, Santa Barbara, 
CA 93106, USA\\
$^{6}$ University of California Riverside, Riverside, CA 92521, USA\\
$^{7}$ School of Physics and Astronomy, University of Birmingham,
Edgbaston, Birmingham B15 2TT
}

\date{Accepted 2003 February 25.  Received 2003 February 10; in original form
2002 June 13}
\pagerange{\pageref{firstpage}--\pageref{lastpage}}
\pubyear{2003}
 
\label{firstpage}

\maketitle
 
\begin{abstract}
We present the results of a {\em Chandra} ACIS-I survey of a high
latitude region at 13h +38 which was earlier observed with {\em ROSAT}
(\mch \etal 1998) and which has recently been observed by {\em
XMM-Newton} for 200ksec. {\em XMM-Newton} will provide good quality
X-ray spectra for over 200 sources with fluxes around the knee of the
logN/logS, which are responsible for the bulk of the X-ray background
(XRB). The main aim of the {\em Chandra} observations is to provide
arcsecond, or better, positions, and hence reliable identifications,
for the {\em XMM-Newton} sources.  The ACIS-I observations were
arranged in a mosaic of four 30ksec pointings, covering almost all of
the $15'$ radius {\em XMM-Newton}/{\em ROSAT} field.  We detect 214
{\em Chandra} sources above a Cash likelihood statistic of 25, which
approximates to $5 \sigma$ significance, to a limiting flux of $\sim
1.3 \times 10^{-15}$ \ecs ($0.5-7$ keV).

Optical counterparts are derived from a Subaru SuprimeCam image
reaching to $R\sim 27$.  The very large majority of the {\em Chandra}
sources have an optical counterpart, with the distribution peaking at
$23<R<24$, although 14 have no counterpart to $R=27$.  The fraction of
X-ray sources with no identification brighter than $R=27$ is similar
to that found in deeper \ch surveys (eg Hornschemeier \etal 2001; see
Alexander \etal 2001 for a detailed discussion of X-ray sources with
faint optical identifications).

The majority of the identifications are with galaxies.  As found in
other {\em Chandra} surveys, there is a very wide range of optical
magnitude for given X-ray flux, implying a range of emission
mechanisms, and many sources have high $L_{X}/L_{opt}$ ratios,
implying absorption at moderate redshift.

Comparison with the earlier {\em ROSAT} survey shows that the accuracy
of the {\em ROSAT} positions agrees very well with the predictions
from simulations in \mch \etal (1998) and that the large majority of
the identifications were correct.

\end{abstract}
 
 
\section{Introduction}

The deepest surveys with {\em ROSAT} (eg Hasinger \etal 1998;
\mch~\etal 1998) have resolved almost all of the soft ($0.5-2$ keV)
X-ray background (XRB). However the bulk of the energy in the XRB lies
at significantly higher energies ($\sim 30$ keV) and so we must
discover which objects dominate the X-ray sky at energies above the
{\em ROSAT} band if we are to properly understand the XRB.  Very deep
surveys with the {\em Chandra} X-ray Observatory (Weisskopf \etal
1996), eg Mushotzky \etal (2000), Giacconi \etal (2001, 2002),
Hornschemeier \etal (2000, 2001), Brandt \etal (2001a, 2001b), have
now resolved almost all of the background in the $0.5-7$ keV band
(hereafter referred to as the medium energy band).  A compilation of
recent and historical measurements by Moretti \etal (2003) conclude
that $94.3^{+7.0}_{-6.7}$ and $88.8^{+7.8}_{-6.6}$ per cent of the
$0.5-2$ and $2-10$ keV XRB is due to discrete sources.  These
observations have revealed that faint X-ray sources are a far from
homogenous population. The {\em Chandra} surveys, for example, contain
a mix of broad line AGN together with very faint ($I\geq 25$) objects
of unknown type (Alexander \etal 2001), galaxies with only narrow
optical emission lines (NELGs) and some bright ($R=18-22$) optically
inactive galaxies (Mushotzky \etal 2000). The major question now is,
what is the nature of these various classes of X-ray source, eg what
is their X-ray emission mechanism? how important is absorption? It is
also of great interest to know if there are further classes of emitter
which are important at energies above the \ch band.

X-ray spectra provide one of the best diagnostics of the X-ray
emission mechanism and so, with {\em XMM-Newton} (Jansen
\etal 2001) we have made a 200ksec observation, using the
EPIC cameras (Lumb \etal 2000), of an area which we observed earlier
with {\em ROSAT} (\mch~\etal 1998).  {\em XMM-Newton} was specifically
designed for spectral investigations and covers the wide band $0.1-12$
keV. At lower energies it has $\sim4 \times$ the throughput of \ch and
its upper energy bound is $\sim4$ keV higher than that of {\em
Chandra}, making it particularly useful for the study of obscured AGN,
much quoted as being the likely major contributors to the XRB (eg
Setti \& Woltjer 1989; Wilman \& Fabian 1999; Gilli \etal 2001).

The centroid of the survey area is at RA 13 34 37.0 Dec +37 54 44
(J2000), in a region of sky of extremely low obscuration ($N_{H} \sim
6.5 \times 10^{19}$ cm$^{-2}$; see \mch~\etal 1998 for details).  It
is very well observed in other wavebands, eg radio (Gunn \etal 2003),
optical (Section 3) and near infrared. In addition, a very deep
mid/far infrared observation will be made with {\em SIRTF} (Fanson
\etal 1998) by Rieke and colleagues.  Preliminary results from our
\xmm observation are presented by Page \etal (2003) and detailed
results will be presented elsewhere (Mason \etal in preparation).

At the faint X-ray flux limits of our {\em XMM-Newton} observation
($\sim 0.5 \times 10^{-15}$ \ecs$\!$, $0.5-2$ keV) we expect many
identifications with very faint ($R>24$) optical counterparts (eg
Giacconi \etal 2001; Hornschemeier \etal 2001).  As the surface
density of such objects is high, we require the best possible X-ray
positions to avoid ambiguity in the identifications.  Although {\em
XMM-Newton} positions are quite reasonable ($\sim 1.5''$ for the
highest significance sources, falling to $\sim3.5''$ at $5 \sigma$
significance, after correction for systematic offsets), they are not
adequate, as we shall quantify in Section \ref{sec:chance}, for
unambiguous optical identification at the flux limit of our
survey. However {\em Chandra} positions for sources close to the
pointing axis have sub-arcsecond accuracy.  We have therefore made
four {\em Chandra} ACIS-I observations of the $15'$ radius {\em
XMM-Newton} field to provide improved positions for the {\em
XMM-Newton} sources.  In the soft ($0.5-2$ keV) band, the {\em
XMM-Newton} source density ($\sim1000$ per square degree above
$5\sigma$ significance in the centre of the field) is close to the
\xmm confusion limit. The {\em Chandra} observations also allow us to
resolve almost all areas of confusion in the {\em XMM-Newton} images.

In this paper we present the {\em Chandra} observations (Section
\ref{sec:chandra}). Our large area coverage compared to
single-pointing \ch surveys, at fluxes around the knee in the medium
energy logN/logS source distribution (Rosati \etal 2002), provides an
excellent sample of the sources which contribute the bulk of the XRB.
Later papers will deal with the astrophysics of the various classes of
sources which contribute to the medium energy XRB, and to the
relationship between those classes and the faint source populations in
other bands, eg radio (cf Gunn \etal 2003). The main aim of the
present paper is to provide the accurate positions and good optical
identifications required for all later investigations.

Excellent optical coverage is provided by a Subaru SuprimeCam image of
almost the whole X-ray area to $R\sim27$ (Section \ref{sec:opt_obs}).
Thus we are able to place stringent limits on the magnitudes of any
{\em Chandra} sources with no optical identification. The
identification content of the survey is discussed in Section
\ref{sec:ids}, including a discussion of the likely contribution of
obscured AGN. The {\em Chandra} observations also enable us to comment
on the accuracy of the positions from our earlier {\em ROSAT} survey,
which we find to be remarkably good, and to clarify which
identifications were correct (Section \ref{sec:rosat}), confirming
many of the narrow emission line galaxies (NELGs).

\section{{\em Chandra} Observations}
\label{sec:chandra}

\subsection{Observational Details}

\begin{figure}
\psfig{figure=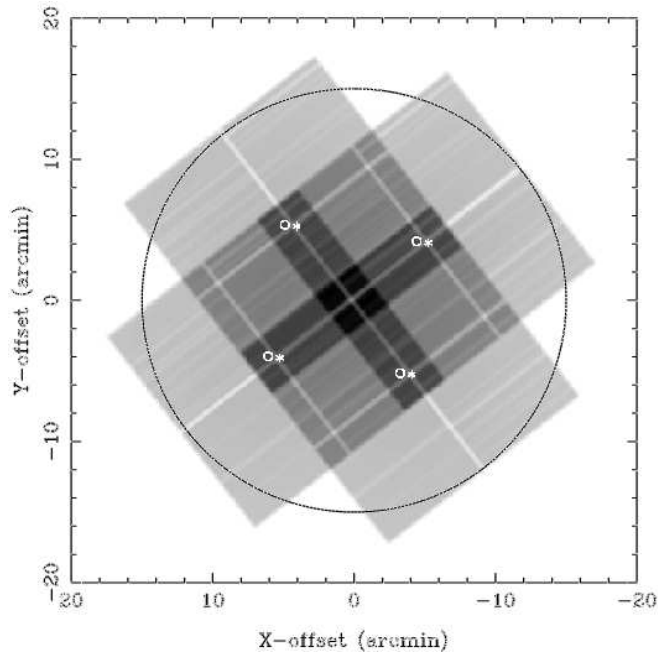,width=3.4in,angle=270}
\caption{Exposure map of the {\em ROSAT}/{\em XMM-Newton} field, which
is marked with the large circle, obtained by our mosaic of 4 {\em
Chandra} pointings. For each pointing the centre of the ACIS-I field
is marked by an asterix and the optical axis of the telescope is
marked by a small circle.}
\label{roughmap}
\end{figure}

In order to cover the large majority of the {\em XMM-Newton} (and {\em
ROSAT}) field at small {\em Chandra} off-axis angles, four pointings
were made with the ACIS-I instrument. The observations were made
consecutively and the roll angle of the observations are identical to
within $0.1''$ providing the coverage pattern shown in
Fig.~\ref{roughmap}. The log of observations is listed in Table
\ref{tab:obslog}. The observations were postponed to avoid a large solar
flare and then took place in relatively low background conditions. We
used the `very faint' observational mode\footnote{\tt 
http://hea-www.harvard.edu/$\sim$alexey/vf\_bg/vfbg.html} in which a
5$\times$5 pixel array is read out around every event, allowing better
photon/cosmic ray discrimination than in the standard 3$\times$3 pixel
`faint source' readout mode.

There is a known positional offset for any {\em Chandra} observation
which depends on the roll angle. For our roll angle that offset is
$\delta$RA=+$1.10''$ and $\delta$Dec=+$1.14''$.  These offsets were
applied to the data before undertaking the source fitting procedure.

\begin{table*}
\centering
\caption {Log of {\em Chandra} observations}
\begin{tabular}{ccccccc}
Field & Sequence &  Observation & \multicolumn{2}{c}{\em Chandra} & Roll   & Exposure \\
Number& Number   &  Date        &  RA     & Dec        & Angle  & (ksec)\\
   1  & 900063 & 8 June 2001 &13 34 13.06 & +37 58 46.0& 232.20 & 30.18\\  
   2  & 900064 & 9 June 2001 &13 35 06.19 & +37 50 34.7& 232.20 & 30.18\\  
   3  & 900065 & 9 June 2001 &13 35 00.43 & +37 59 54.7& 232.20 & 28.53\\      
   4  & 900062 & 8 June 2001 &13 34 18.91 & +37 49 25.9& 232.20 & 30.39   
\end{tabular}
\label{tab:obslog}
\end{table*}

\subsection{Source Searching} 

By comparison with the background spectrum we determined that the band
between 0.5 and 7.0 keV provides the optimum signal/noise for
selection of typical sources and so ACIS images were made, and
searched, in that band. Our source detection code is based on the
algorithm of Cash (1979). The procedure involves a maximum likelihood
process, fitting the PSF of the detector to the photon distribution
and provides the best possible source positions (see Section
\ref{sec:posns}). Details of the procedure are given in \mch \etal
(1998). 

For the present {\em Chandra} observations we have customised the
source fitting software to take account of the variation of PSF with
off-axis angle although, for computational simplicity, we assume a
circular PSF rather than the roughly elliptical PSF which occurs at high
off-axis angles.  We have made a simultaneous fit to the data from all
four ACIS-I pointings, taking account of the different PSFs that may
therefore apply at any particular sky position, to produce a single
probability map of the sky. This procedure makes optimum use of all
data.  Objects are associated with peaks in the map of Cash statistic
over the field.  We set the detection limit at $\chi^{2}>25$ which
corresponds approximately to a $5\sigma$ detection.  The resulting 
catalogue, comprising 214 sources, is given in Table \ref{tab:main}.

\begin{table*}[h] 
\caption{Flux-ordered catalogue of \ch sources and their optical counterparts.}
\noindent
\begin{flushleft}
\begin{minipage}{2.6in}
\begin{tabular}{lp{2.6in}}
Column &  \\
{[1]} & {\em Chandra} flux-ordered source number.\\
{[2]} & Whether there is an alternative optical candidate.\\
 &`U'- the true identification is uncertain.\\ 
 &`P'- the primary identification listed in Table~\ref{tab:main}\\
 & is probably correct. See Appendix A for details. \\
{[3]} &  0.5-7.0 keV flux in units of $10^{-15}$ \ecs \\
{[4]} & Cash $\chi^{2}$ value for the \ch source.\\
{[5,6]} & RA and Dec (J2000) of the \ch centroid.\\
\end{tabular}
\end{minipage}
\hspace{2cm}
\begin{minipage}{2.6in}
\begin{tabular}{lp{2.6in}}
Column &  \\
{[7,8]} & Optical - \ch offset (in that sense), in RA and Dec,
        of the optical counterpart listed in Table~\ref{tab:main}.\\
{[9]} & Total Optical - \ch offset (arcsec).\\
{[10]} & $R$-band magnitude of the optical counterpart.\\
{[11]} & {\em Chandra} off-axis angle of the source (arcmin).\\
{[12]} & {\em XMM} off-axis angle of the source (arcmin).\\
{[13,14]} & FWHM (arcsec) and stellarity of the optical counterpart.\\
\end{tabular}
\end{minipage}
\vspace*{2mm}

Notes to individual sources are at the end of the Table.
\end{flushleft}

\begin{center} 
\begin{tabular}{rcrrrrrrrrrrcc} 
 Num & Alt & Flux & \multicolumn{1}{c}{$\chi^{2}$} & \multicolumn{2}{c}{Chandra} & \multicolumn{1}{c}{$\delta_{O-X}^{\rm RA}$} & \multicolumn{1}{c}{$\delta_{O-X}^{\rm Dec}$} & \multicolumn{1}{c}{$\delta_{O-X}^{\rm Tot}$}  & \multicolumn{1}{c}{R} & \multicolumn{1}{c}{OffC} & \multicolumn{1}{c}{OffX} & FWHM & Stellar \\ 
&&&& \multicolumn{2}{c}{RA (J2000) Dec} & \multicolumn{1}{c}{($''$)} & \multicolumn{1}{c}{($''$)} & \multicolumn{1}{c}{($''$)} & (mag) & \multicolumn{1}{c}{($'$)} & \multicolumn{1}{c}{($'$)} & ($''$) & \\ 
&&&&&&&&&&&&& \\ 
 $^{\P}$1 & & 800.53& 45497.4 & 13:34:51.43 & 37:46:19.10& 0.58 &  0.60 & 0.83 &  13.8 &   5.47 &  8.89 &   0.90 & 1.00 \\  
   2 &   &  85.00 &  1337.7 & 13:33:42.28 & 38:03:35.87 &  1.10 &  0.32 & 1.15 &  18.8 &   7.89 & 13.95 &   1.16 & 1.00 \\  
   3 &   &  80.15 &  3770.6 & 13:34:31.35 & 37:48:31.49 & -0.17 & -0.10 & 0.20 &  20.8 &   2.40 &  6.31 &   0.94 & 0.85 \\  
   4 &   &  69.36 &  1766.2 & 13:33:59.91 & 37:49:12.08 &  0.09 & -0.31 & 0.33 &  17.7 &   4.07 &  9.18 &   1.91 & 0.03 \\  
   5 &   &  67.98 &  3858.3 & 13:34:17.55 & 37:57:22.63 & -0.15 & -0.18 & 0.23 &  18.6 &   1.66 &  4.66 &   2.56 & 1.00 \\  
   6 &   &  63.64 &   928.7 & 13:35:29.77 & 38:04:32.61 & -0.96 & -0.47 & 1.07 &  19.5 &   7.06 & 14.29 &   1.10 & 0.66 \\  
   7 &   &  59.73 &  1485.8 & 13:33:58.55 & 37:59:38.48 &  0.25 & -0.18 & 0.31 &  21.1 &   3.24 &  9.03 &   0.92 & 0.98 \\  
   8 &   &  58.86 &  3775.9 & 13:34:38.08 & 37:56:03.88 &  0.03 &  0.17 & 0.17 &  20.0 &   5.44 &  1.35 &   1.18 & 0.03 \\  
   9 &   &  55.02 &  2821.5 & 13:34:10.63 & 37:59:56.14 &  0.16 &  0.19 & 0.25 &  19.6 &   1.28 &  7.35 &   1.19 & 0.19 \\  
  10 &   &  48.73 &  1972.8 & 13:34:41.83 & 38:00:11.36 &  0.15 & -0.14 & 0.21 &  18.4 &   3.97 &  5.54 &   1.41 & 0.03 \\  
  11 &   &  48.65 &  1912.2 & 13:34:46.94 & 37:47:48.46 &  0.06 &  0.15 & 0.16 &  20.9 &   5.05 &  7.20 &   1.50 & 0.03 \\  
  $^{\dag}$12 &   &  48.22 &   761.3 & 13:35:44.70 & 37:51:40.35 & -0.68 &  0.33 & 0.76 &  20.1 &   7.36 & 13.71 &    --- & ---  \\  
  13 &   &  48.15 &  2519.9 & 13:34:47.37 & 37:59:50.10 &  0.36 & -0.10 & 0.37 &  21.5 &   2.89 &  5.50 &   1.02 & 0.87 \\  
  14 &   &  41.63 &   413.4 & 13:33:32.03 & 37:46:42.06 & -0.15 & -0.86 & 0.88 &  20.4 &  10.00 & 15.14 &   1.01 & 0.98 \\  
  15 &   &  39.60 &   505.9 & 13:35:35.52 & 37:57:45.86 & -0.56 &  0.11 & 0.57 &  20.2 &   7.00 & 11.93 &   0.91 & 0.98 \\  
  16 &   &  36.56 &   432.8 & 13:35:25.45 & 38:05:34.30 & -1.14 & -0.54 & 1.26 &  15.7 &   7.18 & 14.43 &   7.22 & 0.69 \\  
  17 &   &  35.04 &  1096.8 & 13:33:55.81 & 37:52:58.52 &  0.20 &  0.18 & 0.27 &  20.6 &   5.92 &  8.31 &   0.93 & 0.99 \\  
  18 &   &  34.39 &  1579.5 & 13:34:42.77 & 37:59:15.03 &  0.19 & -0.06 & 0.19 &  19.8 &   3.87 &  4.66 &   0.95 & 0.99 \\  
  19 &   &  34.03 &  1359.6 & 13:35:15.92 & 37:52:40.72 & -0.04 &  0.01 & 0.04 &  22.2 &   2.52 &  7.95 &   1.25 & 0.06 \\  
  20 &   &  32.72 &  1652.4 & 13:35:06.21 & 37:49:52.89 & -0.11 & -0.45 & 0.47 &  18.8 &   0.91 &  7.54 &   1.93 & 0.03 \\  
  21 &   &  32.29 &  1539.3 & 13:35:02.85 & 37:49:56.61 & -0.06 & -0.21 & 0.22 &  22.6 &   1.25 &  7.00 &   1.23 & 0.05 \\  
  22 &   &  30.84 &  1480.3 & 13:34:52.17 & 37:57:44.71 & -0.04 & -0.05 & 0.07 &  20.1 &   3.03 &  4.24 &   1.27 & 0.17 \\  
  23 &   &  26.57 &   863.5 & 13:34:01.04 & 37:54:04.97 &  0.09 &  0.02 & 0.10 &  21.6 &   5.54 &  7.12 &   0.95 & 0.98 \\  
  24 &   &  26.50 &   222.3 & 13:33:34.06 & 37:45:47.80 & -0.08 & -0.35 & 0.36 &  20.0 &   9.92 & 15.32 &   0.98 & 0.98 \\  
 $^{\P}$25 & & 26.35 & 546.0 & 13:34:08.76 & 38:03:49.42 & 0.28 &  0.08 & 0.30 &  14.9 &   5.03 & 10.66 &   0.90 & 1.00 \\  
  26 &   &  25.27 &  1287.4 & 13:34:36.40 & 37:55:56.85 &  0.07 &  0.22 & 0.23 &  21.0 &   5.23 &  1.22 &   1.41 & 0.03 \\  
  27 &   &  25.20 &   378.3 & 13:33:42.90 & 37:56:37.43 &  0.44 &  0.76 & 0.88 &  23.0 &   6.66 & 10.83 &   1.07 & 0.35 \\  
  28 &   &  24.83 &  1064.4 & 13:34:08.82 & 37:57:06.89 & -0.13 & -0.10 & 0.17 &  22.7 &   2.14 &  6.04 &   0.94 & 0.96 \\  
  29 &   &  23.67 &   393.0 & 13:33:44.23 & 37:57:52.48 &  0.62 &  0.25 & 0.67 &  20.3 &   6.08 & 10.86 &   1.03 & 0.82 \\  
  30 &   &  22.59 &   271.7 & 13:35:30.35 & 37:57:49.53 & -0.57 &  0.26 & 0.63 &  20.8 &   6.03 & 10.96 &   0.89 & 0.98 \\  
  31 &   &  22.52 &   365.3 & 13:35:24.37 & 37:46:14.56 & -0.68 &  0.17 & 0.70 &  22.2 &   5.57 & 12.64 &   1.28 & 0.55 \\  
  32 &   &  22.44 &   392.9 & 13:35:17.64 & 38:02:47.28 & -0.24 & -0.07 & 0.25 &  21.8 &   4.11 & 11.35 &   1.94 & 0.91 \\  
  33 &   &  21.86 &   187.2 & 13:34:38.51 & 38:06:26.46 &  0.07 &  0.75 & 0.75 &  17.7 &   7.86 & 11.71 &   1.62 & 0.03 \\  
  34 &   &  21.36 &   321.3 & 13:35:12.69 & 37:44:18.70 &  0.31 &  1.02 & 1.07 &  20.8 &   6.50 & 12.59 &   1.45 & 0.03 \\  
  35 &   &  19.77 &   218.9 & 13:35:35.80 & 38:01:49.09 & -1.26 &  0.84 & 1.51 &  22.3 &   6.88 & 13.57 &   1.43 & 0.03 \\  
  36 &   &  19.19 &   419.7 & 13:35:22.83 & 37:49:11.06 &  0.45 &  0.62 & 0.77 &  22.3 &   3.36 & 10.62 &   2.20 & 0.03 \\  
  37 &   &  18.82 &   601.2 & 13:35:15.76 & 37:52:16.93 & -0.02 & -0.01 & 0.02 &  24.3 &   2.21 &  8.03 &   1.57 & 0.03 \\  
  38 &   &  17.52 &   560.0 & 13:34:24.61 & 37:46:15.29 & -0.35 & -0.22 & 0.41 &  20.5 &   3.44 &  8.82 &   1.20 & 0.71 \\  
 $^{\dag}$39 & &17.45&200.7 & 13:35:45.15 & 37:49:57.76 & -0.23 &  0.84 & 0.87 &  18.7 &   7.43 & 14.28 &    --- & ---  \\  
  40 &   &  17.01 &   292.0 & 13:33:48.58 & 37:58:09.01 &  0.13 &  0.02 & 0.13 &  21.6 &   5.19 & 10.14 &   0.98 & 0.97 \\  
  41 &   &  17.01 &   204.7 & 13:34:22.17 & 38:06:21.00 &  1.39 & -0.60 & 1.52 &  21.2 &   7.57 & 11.98 &   0.91 & 0.98 \\  
  42 &   &  16.72 &   305.1 & 13:35:12.54 & 38:02:46.95 & -0.64 & -0.10 & 0.65 &  21.2 &   3.41 & 10.67 &   1.60 & 0.88 \\  
  43 &   &  15.71 &   332.9 & 13:34:02.56 & 37:51:29.53 & -0.16 & -0.20 & 0.26 &  20.5 &   4.01 &  7.53 &   0.92 & 0.98 \\  
  44 &   &  15.57 &   106.1 & 13:33:24.32 & 37:57:45.36 & -0.10 &  0.04 & 0.10 &  20.8 &   9.98 & 14.64 &   1.64 & 0.03 \\  
  45 &   &  15.20 &   439.2 & 13:34:34.03 & 38:00:43.48 &  0.35 & -0.17 & 0.39 &  21.5 &   4.23 &  6.02 &   1.11 & 0.76 \\  
  46 &   &  15.13 &   135.2 & 13:33:43.27 & 37:45:11.12 &  0.56 &  0.76 & 0.95 &  23.6 &   8.57 & 14.28 &   1.24 & 0.76 \\  
  47 & U &  14.99 &   152.3 & 13:33:32.09 & 37:58:15.86 &  0.68 &  0.40 & 0.79 &  24.5 &   8.40 & 13.27 &   2.08 & 0.02 \\  
  48 &   &  14.99 &   272.0 & 13:34:01.16 & 37:46:47.90 &  0.25 & -0.06 & 0.26 &  21.8 &   4.73 & 10.64 &   0.87 & 0.98 \\  
  49 &   &  14.77 &   123.7 & 13:33:34.49 & 37:49:12.13 & -0.07 &  0.30 & 0.31 &  19.8 &   9.09 & 13.53 &   1.00 & 0.98 \\  
\end{tabular} 
\end{center} 
\end{table*} 

\addtocounter{table}{-1} 

\begin{table*} 
\caption{(cont.) Flux-ordered catalogue of \ch sources and their 
optical counterparts.}
\begin{center} 
\begin{tabular}{rcrrrrrrrrrrcc} 
 Num & Alt & Flux & \multicolumn{1}{c}{$\chi^{2}$} & \multicolumn{2}{c}{Chandra} & \multicolumn{1}{c}{$\delta_{O-X}^{\rm RA}$} & \multicolumn{1}{c}{$\delta_{O-X}^{\rm Dec}$} & \multicolumn{1}{c}{$\delta_{O-X}^{\rm Tot}$}  & \multicolumn{1}{c}{R} & \multicolumn{1}{c}{OffC} & \multicolumn{1}{c}{OffX} & FWHM & Stellar \\ 
&&&& \multicolumn{2}{c}{RA (J2000) Dec} & \multicolumn{1}{c}{($''$)} & \multicolumn{1}{c}{($''$)} & \multicolumn{1}{c}{($''$)} & (mag) & \multicolumn{1}{c}{($'$)} & \multicolumn{1}{c}{($'$)} & ($''$) & \\ 
&&&&&&&&&&&&& \\ 
  50 &   &  14.19 &    31.3 & 13:33:29.12 & 38:02:18.01 & --- & --- & --- & $\geq$27.0 &   9.57 & 15.36 &    --- & ---  \\  
  51 &   &  14.12 &   116.2 & 13:33:38.79 & 37:52:17.23 &  0.04 &  0.55 & 0.55 &  24.4 &   8.65 & 11.74 &   1.53 & 0.17 \\  
  52 &   &  14.05 &   406.3 & 13:34:03.09 & 37:53:21.83 &  0.30 &  0.12 & 0.32 &  21.7 &   5.09 &  6.83 &   3.24 & 0.03 \\  
  53 &   &  13.61 &   468.9 & 13:34:45.34 & 37:57:22.65 &  0.26 &  0.15 & 0.30 &  20.5 &   4.25 &  3.11 &   0.89 & 0.98 \\  
  54 &   &  13.54 &    85.9 & 13:34:11.11 & 37:39:46.85 & -1.65 & -0.01 & 1.65 &  21.9 &   9.99 & 15.81 &   1.42 & 0.07 \\  
  55 &   &  13.32 &   403.0 & 13:34:31.24 & 37:49:53.14 & -0.08 & -0.24 & 0.25 &  21.7 &   2.15 &  4.98 &   1.21 & 0.32 \\  
  56 & U &  13.10 &    90.7 & 13:33:30.72 & 37:48:15.50 & -0.20 &  1.08 & 1.10 &  24.1 &   9.91 & 14.61 &   1.15 & 0.73 \\  
  57 & U &  12.96 &    70.5 & 13:33:20.30 & 37:57:44.85 &  1.23 &  1.14 & 1.68 &  21.8 &  10.77 & 15.41 &   1.58 & 0.03 \\  
  58 &   &  12.96 &   407.0 & 13:34:08.54 & 37:54:41.82 & -0.17 &  0.15 & 0.23 &  23.5 &   4.40 &  5.61 &   1.20 & 0.15 \\  
  59 &   &  12.38 &   494.6 & 13:34:20.81 & 37:55:01.01 & -0.26 &  0.13 & 0.29 &  20.8 &   4.10 &  3.21 &   1.43 & 0.03 \\  
  60 &   &  11.58 &   117.2 & 13:34:08.20 & 38:06:28.48 & -1.08 & -0.87 & 1.39 &  23.2 &   7.65 & 13.04 &   5.30 & 0.01 \\  
  61 &   &  11.44 &   158.8 & 13:35:03.71 & 37:44:09.44 &  0.22 &  0.94 & 0.97 &  22.9 &   6.63 & 11.82 &   1.57 & 0.03 \\  
  62 &   &  11.44 &    92.6 & 13:33:35.61 & 37:54:00.36 & --- & --- & --- & $\geq$27.0 &   9.13 & 12.13 &    --- & ---  \\  
  63 &   &  11.00 &   465.4 & 13:34:18.92 & 37:58:57.24 &  0.03 & -0.13 & 0.13 &  19.9 &   0.85 &  5.52 &   1.45 & 0.02 \\ 
  64 &   &  10.86 &   280.7 & 13:34:52.04 & 37:58:25.88 &  0.04 & -0.32 & 0.32 &  19.8 &   2.56 &  4.74 &   1.54 & 0.04 \\  
  65 &   &  10.43 &   446.3 & 13:34:46.33 & 37:54:41.23 &  0.15 &  0.15 & 0.21 &  24.0 &   5.78 &  1.84 &   2.94 & 0.02 \\  
  66 &   &  10.28 &   268.6 & 13:34:11.28 & 37:47:57.51 & -0.11 & -0.42 & 0.44 &  21.8 &   2.44 &  8.47 &   0.89 & 0.98 \\  
  67 &   &   9.92 &   286.8 & 13:35:14.84 & 37:50:36.78 &  0.22 &  0.02 & 0.22 &  24.6 &   1.41 &  8.53 &   2.16 & 0.01 \\  
  68 &   &   9.85 &   331.4 & 13:34:08.29 & 37:52:19.70 & -0.03 &  0.12 & 0.12 &  21.4 &   3.64 &  6.15 &   0.91 & 0.97 \\  
  69 &   &   9.77 &   263.9 & 13:34:31.25 & 38:03:09.86 &  0.40 & -0.18 & 0.44 &  24.9 &   5.36 &  8.51 &   2.47 & 0.08 \\  
  70 &   &   9.77 &   278.7 & 13:35:00.20 & 37:56:32.75 &  0.13 &  0.32 & 0.34 &  19.7 &   3.55 &  4.92 &   2.14 & 0.06 \\  
  71 &   &   9.70 &   392.4 & 13:34:27.29 & 37:50:09.88 &  0.00 & -0.18 & 0.18 &  24.0 &   1.47 &  4.95 &   1.04 & 0.85 \\  
  72 &   &   9.56 &   121.6 & 13:33:53.53 & 38:02:04.52 &  0.48 &  0.39 & 0.62 &  23.5 &   5.21 & 11.28 &   1.21 & 0.16 \\  
  73 &   &   9.19 &    79.2 & 13:33:37.54 & 37:47:57.28 & -0.33 &  0.38 & 0.50 &  21.2 &   8.63 & 13.56 &   1.20 & 0.04 \\  
  74 &   &   9.05 &    34.4 & 13:35:04.50 & 37:39:53.56 &  1.46 &  1.03 & 1.79 &  23.5 &  10.87 & 15.81 &   2.06 & 0.03 \\  
  75 &   &   8.91 &   162.8 & 13:35:14.69 & 37:48:39.31 & -0.02 & -0.41 & 0.41 &  23.5 &   2.50 &  9.61 &   1.81 & 0.03 \\  
  76 &   &   8.91 &   130.0 & 13:33:48.18 & 37:53:33.84 &  0.65 &  0.13 & 0.66 &  21.3 &   7.48 &  9.70 &   1.83 & 0.03 \\  
  77 &   &   8.69 &    81.2 & 13:35:19.35 & 37:43:18.08 & -0.51 &  0.49 & 0.71 &  22.9 &   7.79 & 14.17 &   2.42 & 0.03 \\  
  78 &   &   8.69 &    88.9 & 13:33:46.56 & 38:00:22.24 &  0.14 & -0.10 & 0.17 &  24.3 &   5.71 & 11.42 &   2.52 & 0.01 \\  
  79 &   &   8.69 &    84.2 & 13:33:42.48 & 37:50:27.40 &  0.03 & -0.57 & 0.57 &  23.9 &   7.55 & 11.58 &   1.60 & 0.05 \\  
  80 &   &   8.69 &   273.4 & 13:34:29.97 & 37:56:40.08 &  0.15 &  0.32 & 0.36 &  19.3 &   3.78 &  2.38 &   1.81 & 0.03 \\  
 $^{\P}$81& &8.47 &   350.7 & 13:34:14.71 & 37:51:31.31 &  0.01 & -0.21 & 0.21 &  13.4 &   2.24 &  5.45 &   0.90 & 1.00 \\  
  82 &   &   8.40 &   204.8 & 13:34:08.58 & 37:57:14.95 & -0.38 &  0.76 & 0.85 &  21.5 &   2.06 &  6.14 &   2.34 & 0.03 \\  
  83 &   &   8.25 &   202.6 & 13:35:15.24 & 37:58:38.38 & -0.31 & -0.08 & 0.32 &  20.5 &   2.98 &  8.49 &   0.97 & 0.99 \\  
  84 &   &   8.18 &   149.4 & 13:34:00.87 & 38:01:25.03 & --- & --- & --- & $\geq$27.0 &   3.67 &  9.76 &    --- & ---  \\  
  85 & P &   7.89 &   150.3 & 13:34:13.07 & 37:58:30.96 &  0.05 &  0.21 & 0.22 &  22.9 &   0.51 &  6.05 &   1.20 & 0.07 \\  
  86 &   &   7.24 &   107.4 & 13:35:17.33 & 37:54:15.51 & -0.32 &  0.39 & 0.51 &  23.5 &   3.99 &  7.97 &   1.02 & 0.82 \\  
  87 &   &   7.10 &   168.7 & 13:34:38.26 & 38:01:38.10 & -0.10 &  0.01 & 0.10 &  23.1 &   4.92 &  6.91 &   1.28 & 0.04 \\  
  88 &   &   6.95 &    54.4 & 13:34:13.77 & 38:07:23.05 & -0.13 &  1.07 & 1.08 &  24.0 &   8.46 & 13.45 &   2.00 & 0.03 \\  
  89 &   &   6.95 &    44.6 & 13:33:31.36 & 38:00:48.46 &  0.01 &  0.60 & 0.60 &  23.0 &   8.72 & 14.29 &   1.46 & 0.17 \\  
  90 &   &   6.88 &   172.1 & 13:35:14.78 & 37:52:58.15 & -0.01 & -0.31 & 0.31 &  21.9 &   2.63 &  7.66 &   1.57 & 0.03 \\  
  91 &   &   6.81 &    71.9 & 13:35:29.88 & 37:46:03.13 & -0.32 & -0.03 & 0.32 &  24.0 &   6.41 & 13.59 &   0.95 & 0.96 \\  
  92 &   &   6.81 &    42.8 & 13:34:50.74 & 38:07:05.72 & -0.29 & -1.02 & 1.06 &  22.1 &   7.36 & 12.65 &   1.36 & 0.03 \\  
  93 &   &   6.81 &   179.9 & 13:34:19.93 & 37:54:00.48 & -0.03 &  0.22 & 0.22 &  25.7 &   4.41 &  3.45 &   1.25 & 0.42 \\  
  94 &   &   6.66 &   197.9 & 13:34:35.86 & 37:54:19.04 & -0.37 & -0.23 & 0.44 &  25.4 &   5.62 &  0.47 &   1.22 & 0.64 \\  
  95 & P &   6.59 &   130.6 & 13:34:37.20 & 37:54:36.78 & -0.03 & -0.30 & 0.30 &  23.6 &   6.01 &  0.13 &   1.05 & 0.70 \\  
  96 &   &   6.59 &    56.9 & 13:35:19.24 & 37:42:59.61 & -0.22 &  0.30 & 0.37 &  22.1 &   8.07 & 14.41 &   3.74 & 0.03 \\  
  97 &   &   6.59 &   101.3 & 13:34:36.74 & 38:03:19.00 &  0.57 & -0.12 & 0.58 &  22.6 &   5.93 &  8.58 &   0.97 & 0.98 \\  
  98 &   &   6.37 &    45.8 & 13:35:31.98 & 37:45:35.95 & -0.70 &  0.35 & 0.78 &  17.9 &   7.03 & 14.20 &   2.04 & 0.03 \\  
  $^{\dag}$99 & & 6.23&49.1 & 13:35:44.04 & 37:47:31.89 & -0.36 &  2.78 & 2.80 &  21.7 &   7.86 & 15.08 &    --- & ---  \\  
 100 &   &   6.23 &   121.5 & 13:35:09.75 & 37:48:20.25 & -0.08 & -0.03 & 0.09 &  22.4 &   2.44 &  9.10 &   1.23 & 0.06 \\  
 101 &   &   5.86 &   125.3 & 13:35:20.07 & 37:58:24.11 & -0.64 & -0.01 & 0.64 &  22.6 &   3.94 &  9.25 &   0.98 & 0.98 \\  
 102 & U &   5.79 &    54.0 & 13:35:43.14 & 37:53:08.69 & -0.18 &  0.87 & 0.89 &  23.9 &   7.39 & 13.15 &   1.18 & 0.94 \\  
 103 &   &   5.72 &    68.4 & 13:34:58.38 & 38:04:30.36 & -0.02 & -0.53 & 0.53 &  20.4 &   4.49 & 10.64 &   1.23 & 0.94 \\  
 104 &   &   5.57 &    32.6 & 13:34:28.58 & 37:41:27.74 & -0.44 &  0.85 & 0.96 &  25.3 &   8.29 & 13.38 &   2.76 & 0.11 \\  
 105 &   &   5.57 &    30.7 & 13:33:37.05 & 37:56:29.65 &  0.41 &  1.30 & 1.36 &  20.9 &   7.79 & 11.95 &   0.95 & 0.99 \\  
 106 &   &   5.50 &   161.5 & 13:34:36.24 & 37:51:06.63 & -0.05 & -0.05 & 0.07 &  21.2 &   3.47 &  3.63 &   0.91 & 0.90 \\  
 107 &   &   5.50 &    63.1 & 13:35:35.80 & 37:51:10.81 & -0.24 &  1.12 & 1.15 &  23.3 &   5.56 & 12.14 &   2.10 & 0.03 \\  
 108 &   &   5.36 &   164.9 & 13:34:34.62 & 37:56:39.34 & --- & --- & --- & $\geq$27.0 &   4.56 &  1.98 &    --- & ---  \\  
 109 &   &   5.36 &    26.9 & 13:34:54.02 & 38:07:56.28 &  1.05 & -2.61 & 2.82 &  22.6 &   8.02 & 13.62 &   1.26 & 0.28 \\  
 110 &   &   5.21 &    53.7 & 13:34:24.02 & 37:42:57.83 & -0.21 &  0.37 & 0.43 &  21.5 &   6.67 & 12.05 &   1.41 & 0.03 \\  
 111 &   &   5.21 &    31.4 & 13:33:59.05 & 38:05:56.74 &  1.11 & -1.49 & 1.86 &  22.4 &   7.65 & 13.47 &   1.05 & 0.63 \\  
 112 &   &   5.14 &   157.0 & 13:34:57.28 & 37:49:43.53 &  0.48 & -0.00 & 0.48 &  25.5 &   2.30 &  6.41 &   1.44 & 0.24 \\  
 \end{tabular} 
\end{center} 
\end{table*} 

\addtocounter{table}{-1} 

\begin{table*} 
\caption{(cont.) Flux-ordered catalogue of \ch sources and their 
optical counterparts.}
\begin{center} 
\begin{tabular}{rcrrrrrrrrrrcc} 
 Num & Alt & Flux & \multicolumn{1}{c}{$\chi^{2}$} & \multicolumn{2}{c}{Chandra} & \multicolumn{1}{c}{$\delta_{O-X}^{\rm RA}$} & \multicolumn{1}{c}{$\delta_{O-X}^{\rm Dec}$} & \multicolumn{1}{c}{$\delta_{O-X}^{\rm Tot}$}  & \multicolumn{1}{c}{R} & \multicolumn{1}{c}{OffC} & \multicolumn{1}{c}{OffX} & FWHM & Stellar \\ 
&&&& \multicolumn{2}{c}{RA (J2000) Dec} & \multicolumn{1}{c}{($''$)} & \multicolumn{1}{c}{($''$)} & \multicolumn{1}{c}{($''$)} & (mag) & \multicolumn{1}{c}{($'$)} & \multicolumn{1}{c}{($'$)} & ($''$) & \\ 
&&&&&&&&&&&&& \\ 
 113 &   &   5.14 &    61.5 & 13:33:42.69 & 37:52:39.48 &  0.50 &  0.30 & 0.58 &  23.7 &   8.06 & 10.92 &   1.37 & 0.04 \\  
 $^{\dag}$114 & & 5.07& 34.9& 13:35:52.42 & 37:50:43.74 &  0.00 & -0.22 & 0.22 & $>$24 &   8.82 & 15.42 &    --- & ---  \\  
 115 & U &   5.07 &    60.8 & 13:35:36.41 & 37:51:11.58 & -0.03 &  0.67 & 0.67 &  26.0 &   5.68 & 12.25 &   1.57 & 0.52 \\  
 116 &   &   5.00 &   136.0 & 13:34:19.16 & 37:50:30.21 &  0.10 &  0.22 & 0.24 &  22.0 &   0.94 &  5.50 &   2.30 & 0.03 \\  
 117 &   &   5.00 &    89.0 & 13:34:23.07 & 38:04:15.67 &  0.28 &  0.34 & 0.43 &  19.9 &   5.59 &  9.91 &   1.00 & 0.98 \\  
 118 &   &   5.00 &    29.0 & 13:34:01.90 & 38:08:14.36 &  0.46 &  0.04 & 0.46 &  23.4 &   9.64 & 15.17 &   1.03 & 0.94 \\  
 119 &   &   4.85 &   141.1 & 13:34:14.28 & 37:52:31.60 &  0.06 &  0.57 & 0.57 &  23.5 &   3.17 &  5.00 &   6.16 & 0.03 \\  
 120 &   &   4.85 &    82.9 & 13:34:14.78 & 38:00:01.15 &  0.03 & -0.28 & 0.28 &  23.9 &   1.09 &  6.86 &   1.50 & 0.48 \\  
 121 &   &   4.78 &    54.4 & 13:35:18.04 & 37:46:25.40 & -0.36 &  0.14 & 0.39 &  23.9 &   4.77 & 11.61 &   1.97 & 0.03 \\  
 122 & P &   4.78 &   101.1 & 13:34:28.56 & 37:47:07.05 & -0.52 & -0.35 & 0.63 &  24.0 &   2.95 &  7.80 &   3.58 & 0.26 \\  
 123 &   &   4.78 &    31.5 & 13:35:06.00 & 38:06:59.12 & --- & --- & --- & $\geq$27.0 &   6.96 & 13.51 &    --- & ---  \\  
 124 &   &   4.78 &    51.1 & 13:33:56.42 & 38:04:07.14 &  0.78 & -0.18 & 0.80 &  22.4 &   6.30 & 12.32 &   1.12 & 0.34 \\  
 125 &   &   4.71 &    82.3 & 13:35:19.38 & 37:53:00.25 & -0.47 & -0.28 & 0.55 &  22.1 &   3.23 &  8.54 &   2.03 & 0.03 \\  
 126 &   &   4.71 &    33.1 & 13:35:23.38 & 38:04:41.39 & -0.98 & -0.28 & 1.02 &  19.8 &   6.25 & 13.51 &   0.98 & 0.98 \\ 
 127 &   &   4.63 &   112.7 & 13:34:43.43 & 37:49:21.98 & -0.06 & -0.32 & 0.33 &  23.3 &   4.54 &  5.52 &   2.71 & 0.02 \\  
 128 &   &   4.56 &    47.0 & 13:33:50.50 & 37:49:46.92 &  1.01 &  0.13 & 1.02 &  23.4 &   5.92 & 10.43 &   4.32 & 0.01 \\  
 129 &   &   4.56 &    27.3 & 13:33:50.83 & 38:05:03.32 & --- & --- & --- & $\geq$27.0 &   7.71 & 13.75 &    --- & ---  \\  
 130 & U &   4.56 &    72.0 & 13:34:44.09 & 37:44:34.90 & -0.76 &  0.02 & 0.76 &  24.7 &   6.85 & 10.25 &   2.95 & 0.14 \\  
 131 &   &   4.42 &    73.3 & 13:35:19.69 & 37:49:18.05 & -0.14 &  0.29 & 0.32 &  23.2 &   2.77 & 10.03 &   2.00 & 0.02 \\  
 132 &   &   4.42 &    50.2 & 13:34:14.11 & 38:04:38.43 &  0.48 &  1.42 & 1.50 &  25.3 &   5.71 & 10.88 &   2.96 & 0.03 \\  
 133 &   &   4.42 &    47.1 & 13:34:21.89 & 38:04:50.31 &  0.19 & -0.02 & 0.19 &  22.4 &   6.08 & 10.53 &   1.18 & 0.07 \\  
 134 &   &   4.34 &    32.4 & 13:35:46.50 & 37:53:27.62 & -1.41 &  1.70 & 2.21 &  23.5 &   8.12 & 13.77 &   1.06 & 0.88 \\  
 135 &   &   4.34 &    73.2 & 13:35:05.57 & 37:50:31.22 &  0.21 & -0.38 & 0.44 &  22.4 &   0.48 &  7.04 &   1.23 & 0.11 \\  
 136 &   &   4.34 &    43.6 & 13:33:52.10 & 38:02:52.10 &  0.02 &  1.05 & 1.05 &  24.7 &   5.93 & 12.01 &   2.59 & 0.01 \\  
 137 &   &   4.20 &    46.8 & 13:34:19.40 & 37:43:49.38 & -0.16 &  0.12 & 0.20 &  23.1 &   5.78 & 11.45 &   1.31 & 0.05 \\  
 138 &   &   4.20 &    31.9 & 13:34:57.72 & 38:05:23.58 & --- & --- & --- & $\geq$27.0 &   5.38 & 11.41 &    --- & ---  \\  
 139 &   &   4.13 &    31.0 & 13:35:38.79 & 37:55:09.91 &  0.19 &  0.42 & 0.46 &  23.5 &   7.56 & 12.19 &   2.27 & 0.02 \\  
 140 &   &   4.05 &    45.3 & 13:35:03.68 & 37:45:15.12 &  0.92 &  0.86 & 1.25 &  23.6 &   5.55 & 10.85 &   1.33 & 0.05 \\  
 141 &   &   4.05 &    94.9 & 13:34:09.90 & 37:54:31.78 & -0.08 &  0.28 & 0.29 &  23.3 &   4.50 &  5.35 &   1.59 & 0.03 \\  
 142 &   &   3.98 &    78.1 & 13:34:45.31 & 38:00:30.43 &  0.05 & -0.41 & 0.41 &  24.3 &   3.31 &  6.00 &   0.94 & 0.96 \\  
 143 &   &   3.98 &    29.2 & 13:33:40.73 & 37:52:42.13 &  1.44 &  1.05 & 1.78 &  22.1 &   8.43 & 11.29 &   0.97 & 0.98 \\  
 144 &   &   3.98 &    89.7 & 13:34:33.51 & 37:48:36.04 &  0.00 & -0.02 & 0.02 &  19.5 &   2.76 &  6.17 &   1.41 & 0.05 \\  
 145 &   &   3.84 &    33.5 & 13:35:03.08 & 37:44:07.40 & -1.25 & -1.01 & 1.60 &  23.9 &   6.68 & 11.80 &   4.62 & 0.37 \\  
 146 &   &   3.76 &    73.9 & 13:35:16.34 & 37:56:21.61 & -0.45 & -0.39 & 0.59 &  16.0 &   4.67 &  7.92 &   5.76 & 0.03 \\  
 147 &   &   3.69 &    43.4 & 13:35:15.05 & 38:03:20.02 & -0.62 & -0.23 & 0.66 &  22.0 &   4.15 & 11.40 &   0.92 & 0.90 \\  
 148 &   &   3.62 &    76.5 & 13:35:08.78 & 37:57:04.82 & -0.22 & -0.65 & 0.69 &  23.4 &   3.28 &  6.69 &   1.03 & 0.81 \\  
  $^{*}$149&U& 3.62 &  80.1 & 13:34:01.19 & 37:53:49.14 & -0.60 &  0.16 & 0.61 &  23.8 &   5.68 &  7.12 &   ---  & ---  \\  
 150 &   &   3.62 &    89.8 & 13:34:56.57 & 37:53:49.98 &  0.35 &  0.26 & 0.43 &  24.7 &   3.80 &  3.96 &   1.41 & 0.21 \\  
 151 &   &   3.55 &    48.7 & 13:35:01.20 & 37:59:38.05 & -0.13 & -0.87 & 0.88 &  22.9 &   0.47 &  6.84 &   1.35 & 0.03 \\  
 152 &   &   3.48 &    29.2 & 13:34:39.05 & 37:43:34.94 & -1.36 & -1.05 & 1.72 &  23.2 &   7.05 & 11.16 &   0.87 & 0.98 \\  
 153 &   &   3.48 &    86.0 & 13:34:48.26 & 37:51:10.25 & --- & --- & --- & $\geq$27.0 &   3.87 &  4.20 &    --- & ---  \\  
 154 & P &   3.40 &    39.9 & 13:35:34.89 & 37:50:28.74 &  0.45 &  0.18 & 0.49 &  24.3 &   5.37 & 12.20 &   2.13 & 0.02 \\  
 155 &   &   3.40 &    61.7 & 13:35:18.49 & 37:55:33.21 & -0.46 &  0.19 & 0.49 &  22.9 &   5.26 &  8.22 &   1.12 & 0.98 \\  
 156 &   &   3.40 &    34.6 & 13:35:25.71 & 37:52:34.93 & --- & --- & --- & $\geq$27.0 &   4.00 &  9.85 &    --- & ---  \\  
 157 &   &   3.40 &    59.1 & 13:34:43.90 & 38:02:45.77 &  0.50 & -0.02 & 0.50 &  24.4 &   4.46 &  8.14 &   3.21 & 0.26 \\  
 158 &   &   3.33 &    53.2 & 13:34:22.05 & 38:04:11.14 & -0.03 & -0.14 & 0.14 &  24.9 &   5.46 &  9.90 &   1.36 & 0.24 \\  
 159 &   &   3.26 &    36.2 & 13:35:06.17 & 37:50:03.93 & --- & --- & --- & $\geq$27.0 &   0.74 &  7.41 &    --- & ---  \\  
 160 &   &   3.19 &    77.3 & 13:34:58.83 & 37:50:17.67 & -0.19 & -0.39 & 0.44 &  23.6 &   1.81 &  6.19 &   2.71 & 0.03 \\  
 161 &   &   3.19 &    66.7 & 13:34:35.10 & 37:49:11.03 &  0.09 &  0.20 & 0.22 &  22.6 &   2.92 &  5.56 &   1.55 & 0.03 \\  
 162 & U &   3.11 &    30.6 & 13:34:33.62 & 38:05:40.27 &  0.32 &  0.55 & 0.64 &  23.8 &   7.71 & 10.96 &   2.86 & 0.02 \\  
 163 &   &   3.11 &    28.4 & 13:35:38.29 & 37:52:06.85 & -0.86 &  1.23 & 1.50 &  22.8 &   6.19 & 12.38 &   1.32 & 0.04 \\  
 164 &   &   3.11 &    39.0 & 13:35:19.33 & 38:00:05.56 & -0.15 & -0.17 & 0.23 &  23.8 &   3.42 &  9.91 &   1.12 & 0.32 \\  
 165 &   &   2.97 &    65.4 & 13:34:53.78 & 37:51:08.19 & -0.08 & -0.30 & 0.31 &  26.4 &   2.78 &  4.89 &   1.52 & 0.62 \\  
 166 &   &   2.90 &    33.4 & 13:35:07.80 & 37:45:44.48 & -0.86 & -0.12 & 0.87 &  26.2 &   5.00 & 10.86 &   0.87 & 0.51 \\  
 167 &   &   2.82 &    65.0 & 13:34:46.61 & 37:58:40.29 &  0.52 &  0.14 & 0.54 &  18.6 &   3.34 &  4.37 &   1.58 & 0.03 \\  
 168 &   &   2.82 &    61.1 & 13:34:29.68 & 37:49:19.80 &  0.11 &  0.25 & 0.28 &  23.9 &   1.84 &  5.59 &   2.01 & 0.02 \\  
 169 &   &   2.75 &    26.0 & 13:33:55.01 & 38:03:13.01 &  1.20 &  0.31 & 1.24 &  24.4 &   5.77 & 11.85 &   1.63 & 0.29 \\  
 170 & U &   2.68 &    78.8 & 13:34:17.03 & 37:59:49.48 & -0.02 & -0.02 & 0.03 &  23.3 &   1.02 &  6.43 &   1.04 & 0.97 \\  
 171 &   &   2.68 &    42.0 & 13:34:46.58 & 38:01:25.70 &  0.50 &  0.21 & 0.54 &  23.8 &   3.32 &  6.96 &   1.16 & 0.29 \\  
 172 &   &   2.68 &    50.9 & 13:34:49.80 & 37:54:49.30 & --- & --- & --- & $\geq$27.0 &   5.40 &  2.53 &    --- & ---  \\  
 173 & P &   2.68 &    63.1 & 13:34:42.90 & 37:52:04.09 &  0.36 &  0.29 & 0.47 &  23.2 &   5.07 &  2.91 &   0.91 & 0.92 \\  
 174 &   &   2.61 &    57.5 & 13:34:30.57 & 37:57:02.94 & -0.17 & -0.15 & 0.23 &  22.3 &   3.67 &  2.64 &   1.32 & 0.03 \\  
 175 &   &   2.61 &    31.8 & 13:33:54.72 & 37:51:28.71 &  0.39 & -0.05 & 0.39 &  24.6 &   5.42 &  8.96 &   1.19 & 0.53 \\  
  \end{tabular} 
\end{center} 
\end{table*} 

\addtocounter{table}{-1} 

\begin{table*} 
\caption{(cont.) Flux-ordered catalogue of \ch sources and their 
optical counterparts.}
\begin{center} 
\begin{tabular}{rcrrrrrrrrrrcc} 
 Num & Alt & Flux & \multicolumn{1}{c}{$\chi^{2}$} & \multicolumn{2}{c}{Chandra} & \multicolumn{1}{c}{$\delta_{O-X}^{\rm RA}$} & \multicolumn{1}{c}{$\delta_{O-X}^{\rm Dec}$} & \multicolumn{1}{c}{$\delta_{O-X}^{\rm Tot}$}  & \multicolumn{1}{c}{R} & \multicolumn{1}{c}{OffC} & \multicolumn{1}{c}{OffX} & FWHM & Stellar \\ 
&&&& \multicolumn{2}{c}{RA (J2000) Dec} & \multicolumn{1}{c}{($''$)} & \multicolumn{1}{c}{($''$)} & \multicolumn{1}{c}{($''$)} & (mag) & \multicolumn{1}{c}{($'$)} & \multicolumn{1}{c}{($'$)} & ($''$) & \\ 
&&&&&&&&&&&&& \\ 
 176 &   &   2.61 &    49.9 & 13:34:57.08 & 37:55:41.51 &  0.05 &  0.57 & 0.57 &  25.7 &   4.49 &  4.07 &   1.90 & 0.14 \\  
 177 &   &   2.61 &    53.4 & 13:35:00.09 & 37:53:43.33 &  0.17 &  0.15 & 0.23 &  23.3 &   3.34 &  4.67 &   0.93 & 0.98 \\  
 178 &   &   2.53 &    32.6 & 13:34:56.82 & 37:45:53.50 & -0.31 &  0.41 & 0.51 &  19.5 &   5.31 &  9.67 &   1.73 & 0.03 \\  
 179 &   &   2.46 &    69.5 & 13:34:30.30 & 37:55:25.16 &  0.07 & -0.29 & 0.29 &  22.6 &   4.68 &  1.49 &   1.70 & 0.03 \\  
 180 &   &   2.39 &    51.9 & 13:34:33.26 & 37:55:09.86 & -0.46 &  0.28 & 0.54 &  20.3 &   5.26 &  0.85 &   1.79 & 0.03 \\  
 181 &   &   2.39 &    59.2 & 13:34:38.11 & 37:49:07.71 & -0.66 & -0.00 & 0.66 &  19.6 &   3.52 &  5.61 &   1.12 & 0.90 \\  
 182 &   &   2.32 &    27.1 & 13:35:33.41 & 37:56:15.75 & -0.16 &  0.05 & 0.17 &  21.7 &   7.28 & 11.23 &   1.19 & 0.04 \\  
 183 &   &   2.24 &    36.9 & 13:34:01.27 & 37:55:14.08 & -0.68 &  1.63 & 1.76 &  19.9 &   4.53 &  7.06 &   2.84 & 0.03 \\  
 184 &   &   2.24 &    46.0 & 13:34:11.54 & 37:59:33.09 &  0.10 & -0.01 & 0.10 &  25.4 &   0.87 &  6.96 &   2.30 & 0.12 \\  
 185 &   &   2.24 &    42.6 & 13:34:23.51 & 37:59:30.85 &  0.00 &  0.33 & 0.33 &  22.3 &   1.85 &  5.47 &   1.60 & 0.03 \\  
 186 &   &   2.17 &    40.2 & 13:35:02.89 & 37:56:18.03 & -0.31 &  0.35 & 0.46 &  22.9 &   3.78 &  5.34 &   8.24 & 0.00 \\  
 187 &   &   2.17 &    36.3 & 13:35:01.69 & 37:47:26.02 & -0.02 &  0.30 & 0.30 &  23.6 &   3.52 &  8.78 &   1.33 & 0.05 \\  
 $^{*}$188 &U& 2.10 &  35.5 & 13:34:01.91 & 37:53:25.73 &  0.44 &  0.47 & 0.64 &  23.6 &   5.30 &  7.05 &   ---  & ---  \\  
 189 &   &   2.03 &    71.4 & 13:34:12.68 & 38:00:23.90 &  0.48 &  0.24 & 0.54 &  24.7 &   1.52 &  7.42 &   1.55 & 0.46 \\
 190 &   &   2.03 &    34.4 & 13:34:48.18 & 37:54:32.91 & --- & --- & --- & $\geq$27.0 &   5.42 &  2.21 &    --- & ---  \\  
 191 &   &   1.95 &    45.6 & 13:34:20.98 & 37:59:30.92 & -0.61 &  0.41 & 0.73 &  23.2 &   1.39 &  5.73 &   2.14 & 0.03 \\  
 192 & U &   1.95 &    57.6 & 13:34:21.56 & 37:49:59.82 &  0.47 &  0.05 & 0.48 &  25.9 &   0.46 &  5.63 &   2.08 & 0.07 \\  
 193 &   &   1.95 &    49.3 & 13:34:28.51 & 37:54:48.32 &  0.76 &  0.13 & 0.77 &  23.8 &   4.95 &  1.68 &   2.67 & 0.05 \\  
 194 &   &   1.95 &    29.8 & 13:35:18.90 & 37:54:16.81 & -0.10 &  0.09 & 0.14 &  23.7 &   4.17 &  8.28 &   3.19 & 0.01 \\  
 195 &   &   1.95 &    27.4 & 13:35:13.08 & 37:56:39.55 & --- & --- & --- & $\geq$27.0 &   4.06 &  7.37 &    --- & ---  \\  
 196 &   &   1.88 &    41.6 & 13:34:22.13 & 37:53:44.87 & -0.19 &  0.62 & 0.65 &  24.6 &   4.17 &  3.09 &   1.12 & 0.86 \\  
 197 &   &   1.88 &    29.6 & 13:34:01.21 & 37:55:24.48 &  0.44 &  0.32 & 0.54 &  23.6 &   4.40 &  7.09 &   1.90 & 0.02 \\  
 198 &   &   1.88 &    26.4 & 13:34:28.11 & 37:47:48.00 &  0.36 &  0.02 & 0.36 &  23.7 &   2.35 &  7.15 &   2.38 & 0.02 \\  
 199 &   &   1.81 &    26.9 & 13:34:54.67 & 37:52:38.57 &  0.90 & -0.35 & 0.97 &  26.2 &   3.20 &  4.06 &   0.99 & 0.53 \\  
 $^{*}$200 &U& 1.74 &  41.7 & 13:35:09.62 & 37:50:11.89 &  0.11 & -0.49 & 0.50 &  25.5 &   0.66 &  7.88 &    --- & ---  \\  
 201 &   &   1.74 &    42.9 & 13:34:56.74 & 37:52:17.72 &  0.44 &  0.10 & 0.45 &  24.9 &   2.67 &  4.60 &   1.51 & 0.13 \\  
 202 & P &   1.67 &    27.2 & 13:35:20.44 & 37:55:29.25 & -0.48 &  1.09 & 1.19 &  22.0 &   5.37 &  8.60 &   1.39 & 0.04 \\  
 203 &   &   1.67 &    30.5 & 13:34:34.99 & 37:56:50.03 &  0.13 &  0.25 & 0.28 &  21.2 &   4.53 &  2.14 &   1.22 & 0.04 \\  
 204 & U &   1.67 &    26.7 & 13:34:43.12 & 38:00:20.47 &  0.35 & -0.08 & 0.36 &  25.4 &   3.72 &  5.74 &   1.70 & 0.13 \\  
 205 &   &   1.67 &    27.6 & 13:34:46.76 & 37:55:45.22 &  0.01 &  0.43 & 0.43 &  24.2 &   5.26 &  2.18 &   2.44 & 0.02 \\  
 206 &   &   1.67 &    26.1 & 13:34:44.89 & 38:00:35.78 &  0.58 & -0.54 & 0.79 &  23.0 &   3.41 &  6.07 &   1.43 & 0.03 \\  
 207 &   &   1.59 &    25.4 & 13:34:33.85 & 37:45:13.26 & --- & --- & --- & $\geq$27.0 &   5.11 &  9.53 &    --- & ---  \\  
 208 &   &   1.52 &    30.8 & 13:34:04.20 & 37:56:38.92 & -0.23 & -0.01 & 0.23 &  20.4 &   3.07 &  6.74 &   0.93 & 0.98 \\  
 209 &   &   1.52 &    33.9 & 13:34:33.95 & 37:49:46.45 & -0.00 & -0.24 & 0.24 &  24.8 &   2.67 &  5.00 &   1.59 & 0.20 \\  
 210 & U &   1.38 &    26.0 & 13:34:33.22 & 37:52:22.40 & -0.52 & -1.06 & 1.18 &  23.4 &   3.75 &  2.48 &   1.17 & 0.31 \\  
 211 &   &   1.38 &    33.0 & 13:34:56.50 & 38:01:46.18 & -0.28 &  0.49 & 0.56 &  23.9 &   2.01 &  8.02 &   1.35 & 0.09 \\  
 212 &   &   1.38 &    27.7 & 13:34:35.23 & 37:53:56.92 & -0.07 & -0.74 & 0.75 &  23.7 &   5.24 &  0.86 &   1.05 & 0.84 \\  
 213 &   &   1.38 &    26.5 & 13:34:09.79 & 37:57:44.85 & -0.28 & -0.01 & 0.28 &  23.4 &   1.51 &  6.15 &   1.11 & 0.41 \\  
 214 &   &   1.23 &    25.5 & 13:34:22.22 & 37:48:04.67 & -0.62 & -0.36 & 0.72 &  20.0 &   1.56 &  7.27 &   1.38 & 0.03 \\  
\end{tabular} 
\label{tab:main} 
\end{center} 
{\footnotesize
\begin{flushleft}
NOTES\\
$^{\P}$  Optical position is taken direct from the APM scan of the POSSII plates and magnitude is from UH imaging.\\
$^{\dag}$ Magnitude taken from CFHT image. For source 114
we give the coordinates of a very faint object which is not visible
on the CFHT image but which is visible on a deep B-band image (Croom,
private communication). Sources 99 and 114, whose identification is unreliable because of the lower sensitivity imaging,
are not included in any statistical analysis.\\
$^{*}$ Optical counterpart not deblended properly by SExtractor software
so coordinates and magnitude, but not stellarity or FWHM, calculated manually.
\end{flushleft}
}
\end{table*}

Note that our source searching, which compares photon distributions
with the shape of the PSF, is optimised for the detection of point
sources. Thus extended sources, such as clusters of galaxies or even
extended starburst emission in galaxies, may be missed or be given
incorrect fluxes. 
 
Examination of the Cash-statistic map, as well as the raw photon
distributions, shows that positions are good, statistically, to better
than $1''$ in the large majority of cases.  By examining the
X-ray/optical offsets of bright optical counterparts for which the
likelihood of chance random associations is very low, we can obtain a
retrospective view of the X-ray positional accuracy. This
investigation is given in Section \ref{sec:posns}.

\subsection{X-ray Positional Accuracy}
\label{sec:singlepos}

From simulations of a 300ksec ACIS-I observation using the MARX
simulator (Wise \etal 2000) Tozzi \etal (2001) present (their Fig 1) a
distribution of differences between input and output coordinates,
drawn from sources on all parts of the ACIS-I detector. Most sources
appear to have positional errors of $<1$arcsec but Tozzi \etal do not
distinguish sources of different off-axis angles. Giacconi \etal
(2002) determine an average rms positional offset of $0.5''$ between
optical and X-ray coordinates for X-ray sources from a 942ksec ACIS-I
observation and assume an analytic form for the increase in X-ray PSF
with off-axis angle. Hornschemeier \etal (2001), in a 222ksec ACIS-I
exposure, adopt a positional error of $0.5''$ but note that, for
sources $>3'$ off-axis, positions will be worse. We are unaware of any
study of positional errors in relatively short ACIS-I exposures, such
as ours and so in this paper we consider the errors in some detail.
In this section we address how the error changes as a function of
off-axis angle by considering the relative positions of X-ray sources
as detected in different observations. Then, in
Section~\ref{sec:posns}, we derive X-ray positional errors by
reference to bright optical counterparts.

Sixty seven sources appear on more than one pointing. There are 158
positional measurements for these 67 sources. We can obtain an idea of
our positional accuracy as a function of off-axis angle by examining
the differences between the positions derived from single observations
and those derived from the simultaneous fit to all observations, which
are listed in Table~\ref{tab:main}.  In Table \ref{tab:singlepos} we
give the offsets within which 50 per cent and 90 per cent of the
sources lie. We also list the number of detections, N, contributing to
each off-axis bin.  The simultaneous fit position is dominated by the
very accurate positions obtained close to the axis and so we do not
list the offsets for sources with off-axis angles less than $5'$. The
positional accuracy of these sources is considered in
Section~\ref{sec:posns}.  As the four individual pointings were placed
on the corners of a square of side $\sim10'$, a detection at, say,
$8'$ off-axis in one observation tends to imply a detection at $\ltsim
4'$ in another.  In other words we will be able to compare the $8'$
off-axis position with a highly accurate, close to the axis, position
and so obtain a good estimate of the off-axis positional uncertainty.

This method of off-axis positional determination, although based on
only 67 sources, depends only on {\em Chandra} observations and so does not
rely on any assumptions about the accuracy of the optical
identifications. It therefore provides a useful complement to the
positional determination in Section~\ref{sec:posns}. Note however
that, in Table~\ref{tab:singlepos}, we neglect any contribution to the
off-axis error from the error in on-axis positions. The on-axis
positional error is, however, very small ($<0.3''$) and so, when added
in quadrature, adds very little to the off-axis error.

\begin{table}
\centering
\caption{Single pointing offsets }
\begin{tabular}{cccc}
Off-axis & N & 50\% & 90\% \\
arcmin& &arcsec& arcsec\\
&&&\\
6-7 & 23 & 0.65 & 1.40 \\
7-8 & 27 & 0.68 & 1.60 \\
8-9 & 13 & 0.85 & 2.30 \\
$>9$& 15 & 1.15 & 2.75 \\
\end{tabular}
\label{tab:singlepos}
\end{table}

Of the 50 single pointing positions out to $8'$ off-axis, only one
lies $>1.7''$ from the simultaneous fit position and most are much
closer.  Even beyond $9'$ off-axis, only 4 of the 15 offsets are
$>1.3''$.

\subsection{X-ray Flux Determination}

Fluxes are determined from the Maximum Likelihood (ML) parameters of
the Cash statistic, where count-rate is a free parameter. These count
rates automatically compensate for the variation in the PSF and the
complicated exposure map of the observations.  The ML count rates are
then converted to $0.5-7$ keV fluxes by assuming a power-law spectrum
with an energy spectral index, $\alpha = 0.4$, and Galactic absorption
of $N_{H}=6.5 \times 10^{19}$~cm$^{-2}$ (Stark \etal 1992;
Branduardi-Raymont \etal 1994). The $\alpha = 0.4$ spectrum is chosen
because it is similar to that of the diffuse background (eg Gendreau
\etal 1995).  For the majority of sources, this spectral assumption is
the dominant source of error ($\alpha = 0$ would increase the flux by
24 per cent, and $\alpha = 1$ would decrease it by a similar
amount). For the faintest sources, however, the excellent on-axis PSF
means that as few as 6 photons can give rise to a significant
detection, and the flux is then dominated by Poisson noise, giving
errors of up to 40 per cent.

\section{Optical Observations}
\label{sec:opt_obs}

\subsection{CCD Photometry}
\label{sec:ccdphot}

\begin{figure}
\psfig{figure=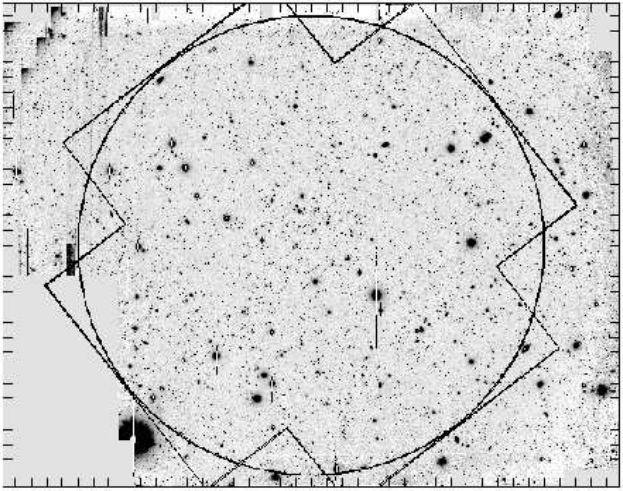,width=3.4in,angle=270}
\caption{Subaru SuprimeCam $R$-band image of the 13hr field, with the
{\em ROSAT}/{\em XMM-Newton} $15'$ radius field-of-view superimposed
with a large circle, as in Fig.~1.}
\label{overview}
\end{figure}

The field was observed for one hour on 2000 December 23 using the
Prime Focus Camera, SuprimeCam (Miyazaki \etal 1998), on the 8.2m
Japanese Subaru Telescope. SuprimeCam consists of ten closely butted
MIT/Lincoln Labs 4096x2048 CCDs, with a scale of 0.2 arcsec/pixel,
giving an overall field of view of $34' \times 27'$. The present
observations were dithered in order to fill the small gaps between the
CCDs, resulting in an overall field of $39' \times 31'$.  At the time
of observation, only nine of the CCDs were operational, hence part of
the south-east corner of the X-ray field was not covered.  An overview
of the optical image is shown in Fig.~\ref{overview}, with a 15 arcmin
radius circle representing the {\em ROSAT} and {\em XMM-Newton} field
overlaid. The limiting magnitude of the observations, for isolated
stellar sources, is $R\sim27$ ($5\sigma$ significance), a little
fainter in some places, although our optical catalogue (see Section
\ref{sec:complete}) is not complete at that limit.  The average seeing
was 0.9 arcsec.

The missing portion of the {\em XMM-Newton} field, containing \ch sources 12,
39, 99 and 114, is covered by an observation from the 8k$\times$8k CCD
camera on CFHT, reaching to $R=24$. For \ch sources 1, 25 and 81, with
very bright stellar counterparts, we show our least saturated optical
imaging from the University of Hawaii (UH) 88 inch telescope, which
reaches to $R=23$ (see \mch \etal 1998 for details).

\subsubsection{Registration of the Subaru Coordinate Frame}

The coordinate frame of the CCD observations was set initially by
reference to the USNO-A2.0 stars (Monet \etal 1998) and also to the
HST Guide Star Catalog (Ver 1.2; van Altena 1999), providing a
coordinate frame which was good to better than an arcsecond. The
coordinate frame was then refined by comparison with stellar objects
visible on the second epoch Palomar Sky Survey J plates. An initial
comparison with scans of the first epoch Palomar plates had shown
evidence of proper motion in significant numbers of stars. The second
epoch plates were scanned by Mike Irwin using the Cambridge Automatic
Plate Measuring Machine, whose frame is very accurately tied to the
FK5 system. The positions of CCD counterparts to the positions of APM
stars were selected automatically (using the {\sc IRAF} task {\sc
imcentroid}) but correspondences were then checked by eye to ensure
that {\sc imcentroid} had not selected a spurious source. Remaining
incorrect associations were removed by an iterative 2.5$\sigma$
cut. Finally over 1100 corresponding objects were used to define the
CCD frame resulting in rms errors of 0.30 arcsec in both RA and Dec
(an improvement of $\sim 0.1''$ in both axes compared to using the
first epoch Palomar plates.) Subsequent comparison of the resultant
CCD positions with those from $0.1''$ accuracy VLA radio positions
(Seymour \etalc, in preparation) confirms the accuracy of the CCD
positions.

\subsection{Object Extraction}

In order to determine the magnitudes, positions and approximate
morphologies of all of the optical objects in the SuprimeCam image,
both to find counterparts to X-ray sources and to determine the
background star and galaxy surface density, we ran the {\sc
SExtractor} software (Bertin \& Arnouts 1996, hereafter BA96).  We ran
{\sc SExtractor} twice: once on the central 6000$\times$4000 pixels
($20.2' \times 13.4'$) of our image where the exposure is very uniform
in order to determine galaxy and star source counts and secondly on
the entire image to find all possible X-ray counterparts.  The input
seeing parameter was set to $0.9''$. Amongst the parameters output by
{\sc SExtractor} is the stellarity parameter, which indicates how
confident the software is that the object detected is a `star' or a
galaxy. Typically, bright objects with stellarity $> 0.5$ are stars or
QSOs and objects with stellarity $<0.5$ are galaxies.  We take the
{\sc SExtractor} `best estimate' magnitude as our standard magnitude.

\begin{figure}
\psfig{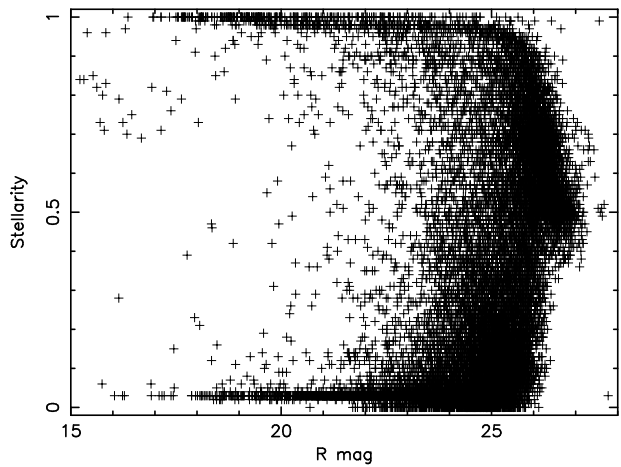}
\caption{Stellarity v. magnitude for objects selected by {\sc SExtractor}
from the central 6000$\times$4000 pixels ($20.2' \times 13.4'$) of the
Subaru $R$-band image.  Stellarity varies between 0 for a galaxy to 1
for a star.}
\label{stellarity}
\end{figure}

In Fig.~\ref{stellarity} we plot the distribution of stellarities for
the central $20.2' \times 13.4'$ as a function of magnitude. The
distribution is typical of the extragalactic sky (eg see fig.~9 of
BA96).  By comparison with the simulations of BA96, and using their
figs 7 and 9, we estimate that almost all objects brighter than $R
\sim 23$ should be correctly classified and that $\sim 80$ per cent of
objects with $23<R<25$ should be correctly classified. Beyond that
magnitude the large majority of objects in the high latitude
extragalactic sky should be galaxies (Smail \etal 1995) but {\sc
SExtractor} cannot give a reliable classification. As will be noted
from BA96 fig.~9 and from our Fig.~\ref{stellarity}, there is a
tendency for {\sc SExtractor} to classify, as stars, faint objects
which are probably galaxies. Thus, for objects fainter than
$R\sim24-25$, imposing a stellarity selection criterion as low as 0.5
probably classifies incorrectly some galaxies as stars. We note that
Hogg \etal (1997) classify all faint objects as galaxies.

\subsection{Background galaxy and `star' counts}
\label{sec:complete}

In order to compare our object extraction efficiency and photometry
with that of previous workers we have calculated the differential
source counts for the objects in our {\sc SExtractor} catalogue.  We
select objects with stellarity $\leq 0.5$. Fitting the distribution in
the range $20.0 < R \leq 25.0$ we find a slope of 0.345 which is
almost identical to the slope determined by Hogg \etal of 0.334 over
the same magnitude range. Taking account of the slight differences
between our Cousins $R$-band and the Johnson $R$-band of Hogg \etal
($R_{J} - R_{C} \sim -0.1$; Fukugita \etal 1995) we find an almost
identical normalisation. At brighter magnitudes we also see the fall
off in galaxies which was noted in the large area photographic surveys
(eg Metcalfe \etal 1991).  Thus our galaxy catalogue appears complete
to $R=25$. Compact galaxies and QSOs, in isolated environments, can be
seen down to $R=27$, but those close to brighter objects are sometimes
not deblended by {\sc SExtractor} so our catalogue is imcomplete at
$R>25$.

We note that the raw galaxy counts of Hogg \etal show exactly the same
deviation from a powerlaw at $R=25$ as do our own counts.  Hogg \etal
correct for incompleteness and show that, to $R=27$, the same powerlaw
that describes the brighter galaxy counts continues to apply. We may
therefore use an extrapolation of the powerlaw which describes our own
integral source counts at $R\leq25$ to estimate the likelihood of a
chance identification, in an optically uncrowded region, in our {\em
Chandra} errorcircles (Section 4.1).

\begin{figure}
\psfig{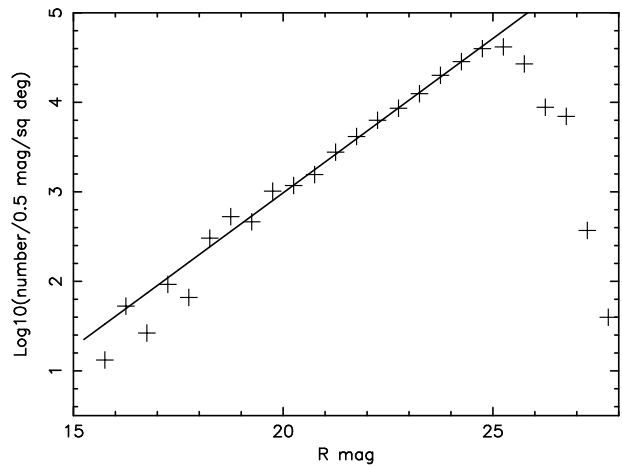}
\caption{Differential number counts, in half magnitude bins, for objects
with stellarity $\leq 0.5$ (ie galaxies). Note that no galaxies were
detected in the first magnitude bin, 15.0 -- 15.5. The line represents
the best fit to the galaxy distribution in the range $20.0 < R \leq
25.0$.  The slope is 0.345. }
\label{logn}
\end{figure}

\section{The Optical Identifications}
\label{sec:ids}

\begin{figure*}
\vspace*{22.5cm}
\caption{$10''\times10''$ $R$-band  Subaru SuprimeCam images centred on
the {\em Chandra} sources. The {\em Chandra} source number is at the
top left of each image, the off-axis angle is in the centre and the
likelihood value is at the top right. The images of {\em Chandra}
sources 12 and 39 are from the CFHT observations and the images of
{\em Chandra} sources 1 and 25 are from the UH 88inch observations. 
All SuprimeCam images have the same greyscale except for the image of
\ch source 16 which is scaled to show brighter features.
 }
\label{rpics1}
\end{figure*}

\addtocounter{figure}{-1} 

\begin{figure*}
\vspace*{22.5cm}
\caption{(cont.) $10''\times10''$ $R$-band  Subaru SuprimeCam images
centred on the {\em Chandra} sources. The {\em Chandra} so urce number
is at the top left of each image, the off-axis angle is in the centre
and the likelihood value is at the top right.  The images of {\em
Chandra} sources 99 and 114 are from the CFHT observations and the
images of {\em Chandra} source 81 is from the UH 88inch observations.
All SuprimeCam images have the same greyscale except for the image of
\ch source 146 which is scaled to show brighter features.}
\label{rpics2}
\end{figure*}

\addtocounter{figure}{-1} 

\begin{figure*}
\vspace*{15.75cm}
\caption{(cont.) $10''\times10''$ $R$-band  Subaru SuprimeCam images
centred on the {\em Chandra} sources. The {\em Chandra} source number
is at the top left of each image, the off-axis angle is in the centre
and the likelihood value is at the top right.  }
\label{rpics3}
\end{figure*}

In order to give as informative a view as possible of the optical
identifications, we show $10''\times 10''$ postage stamp images of all
of the {\em Chandra} sources in Fig.~\ref{rpics1}, in order of \ch
flux.  Unless indicated otherwise in the figure caption, the images
are taken from the Subaru $R$-band image with identical greyscales, to
aid comparison.  All postage stamp images are centred on the {\em
Chandra} positions to within a pixel ($0.2''$).  As the {\em Chandra}
positional accuracy is related to the {\em Chandra} off-axis angle and
the significance of the source detection, each image is labelled with
its {\em Chandra} source number (top left), the smallest {\em Chandra}
off-axis angle from the various {\em Chandra} fields in which the
source was detected (top centre) and the Cash statistic derived in the
source detection (top right).  Visual examination of Fig.~\ref{rpics1}
shows that an object is visible within an arcsecond of the X-ray
position in the very large majority of cases.  However, given the high
accuracy of the {\em Chandra} positions, it is clear that there are a
significant number of sources (28, 13 per cent of our sample) where
the identification is very faint ($R>25$), as has been noted in other
{\em Chandra} surveys (eg Mushotzky \etal 2000; Alexander \etal
2001). Of the 28 sources, 14 have no identification at all to
$R\sim27$.

In Table~\ref{tab:main} we list the {\em Chandra} sources, together
with identification information from the {\sc SExtractor} analysis,
for optical candidates within $2''$ of the \ch centroid.  As we have
seen above (Section~\ref{sec:singlepos}), this selection radius is
sufficient for the vast majority of \ch sources. However $\sim5$ of
the 24 sources at off-axis angles $>8'$ may have identifications in
the range $2-3''$.  For sources more than $7'$ off-axis we therefore
increased the search radius to $3''$. In 2 cases (110 and 135, both
$>8'$ off-axis) a plausible counterpart was found where one had not
already been found within $2''$ of the \ch position. These two
counterparts are listed in Table~\ref{tab:main}.  Examination of the
finding charts for sources $>8'$ off-axis occasionally shows objects
at very large distances ($>3''$) from the \ch position where otherwise
there is no detectable counterpart. Although perhaps one such objects
might be the correct identification, it is very unlikely that many
counterparts will lie so far from the \ch position.

Note that, despite our best efforts, the faintest objects, which are
still easily detected by {\sc SExtractor} and which are clearly
visible on computer displays, are not always clearly visible on the
hard copies presented here.  

\subsection{Positional Accuracy}
\label{sec:posns}

Positional errors for \ch sources are often determined by reference to
unambiguous optical identifications (eg Giacconi \etal 2002;
Hornschemeier \etal 2001). In this section we apply the same method to
our observations. We perform our analysis using X-ray sources with
optical counterparts in the magnitude range $R=19$ to 24 which lie
within $2''$ of the {\em Chandra} position. We choose an $R = 19$
cut-off as stars brighter than $R\sim19$ are saturated on the
SuprimeCam image.  At magnitudes brighter than $R=24$ we will show
below (Section~\ref{sec:chance}) that the number of chance
coincidences is very low.

\begin{figure}
\psfig{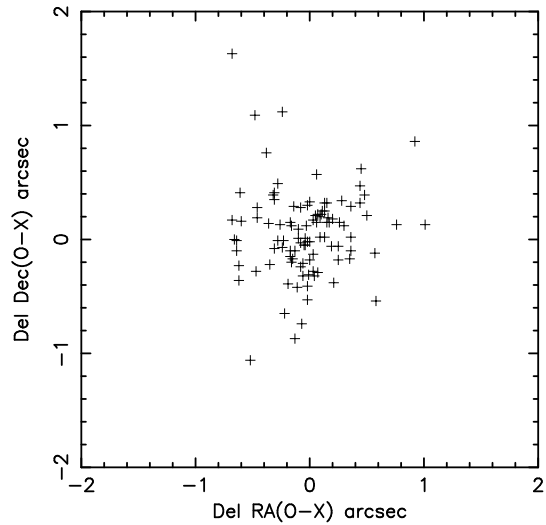}
\caption{Optical/X-ray offsets for all identifications in the range
19 to 24 mag, from \ch sources within 6 arcmin of a \ch pointing axis.}
\label{fig:choffsets}
\end{figure}

We separately calculated the systematic optical/X-ray positional
offsets for all four {\em Chandra} pointings.  The offsets were
reasonably well represented by Gaussian distributions in both right
ascension and declination. In all cases there was evidence of a slight
systematic offset between the optical and X-ray positions but the
offsets were the same, within the errors, for all 4 {\em Chandra}
pointings at $+0.17''$ in RA (optical -- X-ray position) and $+0.06''$
in declination. These small corrections were applied to the {\em
Chandra} coordinate frames and the resultant distribution of
optical-X-ray offsets for all {\em Chandra} sources within 6 arcmin of
a pointing axis and with optical counterparts in the range $19<R<24$
are shown in Fig.~\ref{fig:choffsets}.  For these sources the rms
dispersion between the optical and X-ray coordinates, $\sigma, =
0.28''$ (in both RA and Dec) and for sources further than 6 arcmin
from an axis, going out to $11'$ in one case, $\sigma = 0.80''$. The
average for all sources is $\sigma = 0.40''$.  A full 2D polynomial
coordinate solution between the X-ray and optical positions did not
reduce the dispersion significantly more than the simple systematic
offsets derived above.

In order to investigate the variation of positional accuracy with
off-axis angle we plot, in Fig.~\ref{fig:offax_offsets}, the {\em
Chandra}-optical offset for sources with an identification in the
range $19<R<24$ mag. The lowest significance sources are plotted
separately but off-axis angle is clearly the major factor affecting
source positions.  We conclude that for sources within $6'$ of a {\em
Chandra} pointing axis, the {\em Chandra} positions are good to
typically $\sim 0.7''$, with almost all identifications lying within
$1''$. In the outer part of our \ch images these positional errors
approximately double.

Confirmation of the \ch positions, and the accurate registration of
the \ch and Subaru coordinate frames, can be obtained by examining the
\ch - Subaru offsets for the 28 \ch sources, lying within $6'$ of a
{\em Chandra} pointing axis, which already have firm identifications
(eg \mch \etal 1998). Apart from source 1, whose image is saturated,
none of the other 27 have \ch - Subaru separations greater than
$0.55''$. Of course these 27 are mostly rather bright X-ray sources
and so have the best \ch positions and so reliable identifications can
still be expected, for the fainter \ch sources, at slightly greater
\ch - optical separations.

\subsection{Chance Coincidences}
\label{sec:chance}

\begin{figure}
\psfig{figure=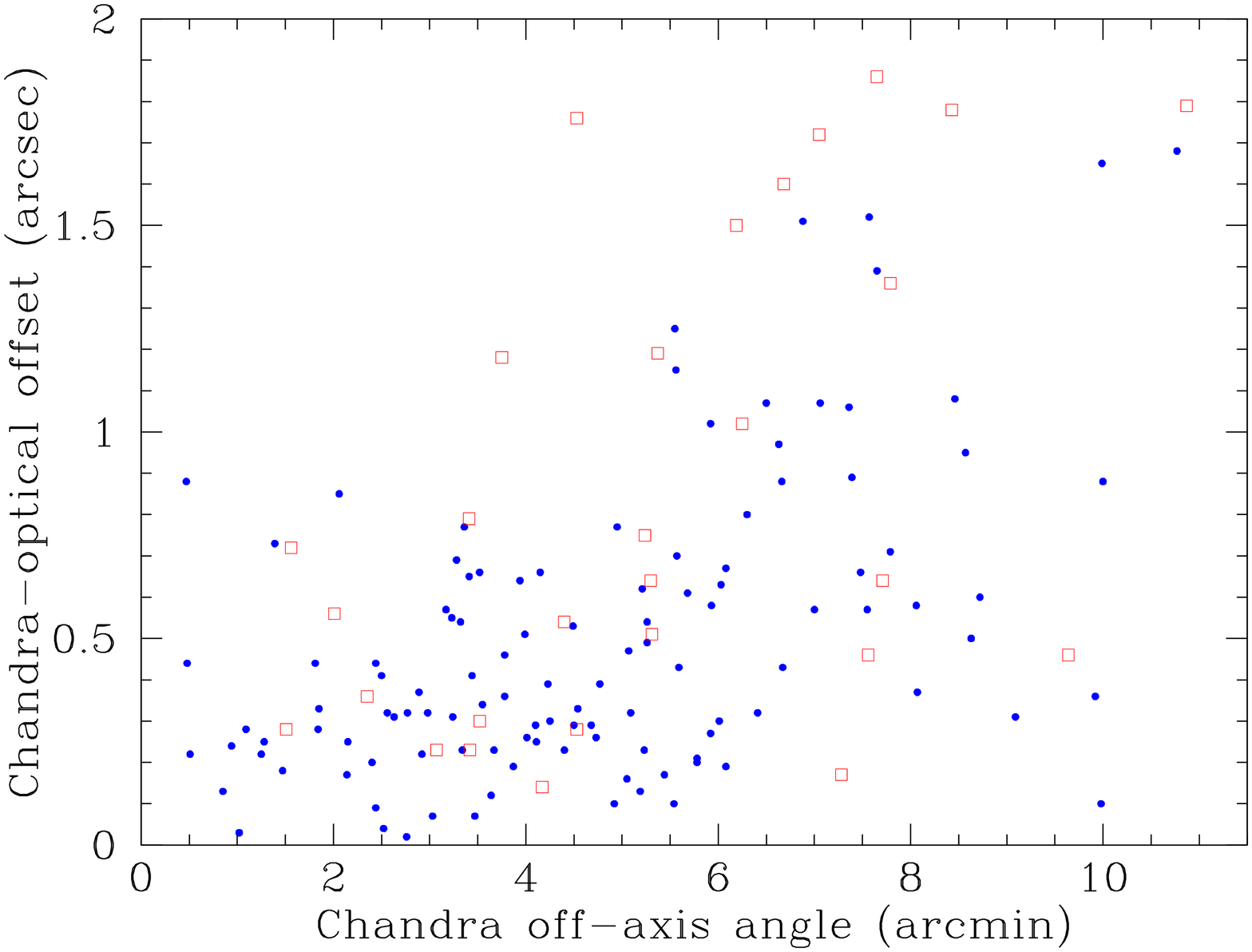,width=3.2in,angle=270}
\caption{Distribution of X-ray/optical offsets as a function of
off-axis angle for the \ch sources with identifications in the
magnitude range $19<R\leq24$. Sources with a detection $\chi^{2} < 40$
are shown as open squares.  }
\label{fig:offax_offsets}
\end{figure}

From examination of Fig.~\ref{fig:offax_offsets} we take $0.7''$ as a
conservative estimate of the 90 per cent confidence radius for sources
within $6'$ of a {\em Chandra} pointing axis. The number of chance
optical coincidences, for the 123 \ch sources with no previous
identification, together with the number of actual identifications, is
given in Table~\ref{tab:chanceids}, for various magnitude ranges. We
see immediately that identifications with $R<25$ can be considered
quite firm, those with $26<R<25$ are mostly reliable, but those with
$R>26$ are probably spurious. We note, however, that the `observed'
numbers include all identifications within our $2''$ search radius. Of
these identifications, 10 lie in the range $0.71 - 1.00''$, 7 lie in
the range $1.01 - 1.25''$ and 2 lie in the range $1.26 - 2.00''$. If
the 90 per cent confidence radius is $0.7''$, then we would expect of
order 10 identifications outside that radius (for the observed 104
within) and so the 10 sources in the range $0.71 - 1.00''$ are
probably good identifications.  Even if we increase the confidence
radius to $1.0''$ and so double the number of chance expected
coincidences, the identifications with $R<25$ are still secure. Of
course, if any faint ($R>25$) identifications are incorrect, then the
true identifications must be even fainter.

Beyond $1.0''$ separation, we are not confident of the
identifications, except for the optically brightest (source 202), even
though most of the larger separations correspond to fainter \ch
sources where poorer positions are expected. If we increase the
confidence radius from $0.7''$ to $1.25''$, we increase chance
coincidences by $\times 3.2$. Thus, to $R<24$, we expect 3.6 chance
coincidences in the 123 \ch errorboxes and there are 5 possible
identifications with $R<24$ in the separation range 1.01 to
$1.25''$. A consequence of our positional accuracy is that we can be
reasonably confident that \ch source 183 is not associated with the
nucleus of the 19.9 magnitude galaxy which lies $1.76''$ away. The
X-ray source must either be associated with a background object, which
is obscured by the galaxy, or with a non-nuclear source (eg an ultra
high luminosity X-ray source (ULX), c.f.~Roberts \& Warwick 2000) in
the outer envelope of the bright galaxy, as may also be the case in
source 188 (see Appendix A for details.)

For \ch sources at greater off-axis angles than $6'$, the positional
errors increase (eg see Fig.~\ref{fig:offax_offsets} and Table
\ref{tab:singlepos}). In the range $6'-8'$ off-axis there are 22
\ch sources with identifications of $R<24$ compared with 2 expected by
chance for a $2''$ radius errorbox, which is sufficient to encompass
all candidates, although larger than the 90 per cent confidence radius
($1.6''$, see Table~\ref{tab:singlepos}).  Thus most of the
identifications with $R<24$ are secure. There are only 3
identifications at fainter magnitudes but those cannot be considered
reliable. Beyond $8'$ off-axis there are 21 \ch sources without
previous identifications, of which 15 have $R<24$. All but 2 of the
latter are within $2''$ of the \ch position.  However taking a
conservative error radius of $3''$ (a little larger than the 90 per
cent confidence radius), we expect 2.7 chance identifications of
$R<24$. Therefore we are again confident that identifications with
$R<24$ are secure but those at fainter magnitudes are not.

\begin{table}
\centering
\caption{Number of sources observed and number of chance coincidences 
expected within $6'$ of a \ch axis.}
\begin{tabular}{ccc}
$R$-magnitude & Observed & Chance \\
&\\
$\leq$24  & 91 & 1.1 \\
 24--25   & 19 & 1.1  \\
 25--26   &  9 & 2.2  \\ 
 26--27   &  4 & 4.5  \\ 
\end{tabular}
\centering
\label{tab:chanceids}
\end{table}

\section{Identification Content}

A number of papers have already been written on the optical
identification of sources in the very deep \ch surveys (eg
Hornschemeier \etal 2001, 2002; Bauer \etal 2002; Tozzi \etal 2001;
Koekemoer \etal 2002; Barger \etal 2002; Vignali \etal 2002; Gunn
\etal 2003).  In this section we summarise the optical properties of
the identifications in our shallower, but larger area, survey to place
them in the context of the deeper surveys so that they may be useful
to other researchers. We also illustrate the usefulness of the {\sc
SExtractor} stellarity parameter for morphological classification.

If our sources lie in broadly the same redshift range as those of
other \ch surveys, ie $\sim 0.1-1.5$ (eg Rosati \etal 2002; Crawford
\etal 2002; Barger \etal 2001) and the redshifts which we have already
(see Table~\ref{tab:rosat}) are consistent with that view, then most
of our sources will be of relatively low luminosity. For example a
flux of $2\times 10^{-15}$\ecs ($0.5-7$ keV), close to the limit of
our catalogue, corresponds to $L_{X} \sim2 \times 10^{40}$ ergs
s$^{-1}$ at redshift $z=0.2$, or $\sim 1.4 \times 10^{43}$ ergs
s$^{-1}$ at $z=1.5$.\footnote{We assume a flat cosmology with
$\Omega_{\rm matter}=0.3$ and $H_{0} = 70$ km s$^{-1}$ Mpc$^{-1}$
throughout this paper.}  Although AGN will certainly be the dominant
sources of X-ray emission, at the lowest luminosities, off-nuclear
emission such as a ULX (eg Roberts and Warwick 2000) is a realistic
possibility (eg sources 183 and 188), as is thermal emission from hot
gaseous halos of large elliptical galaxies (eg O'Sullivan \etal 2001;
Mathews \& Brighenti 1998), starburst emission (Griffiths \& Padovani
1990), and hidden BL Lacs (Browne \& Marcha 1993; \mch \etal 2003).

\subsection{Morphology}

\begin{figure}
\psfig{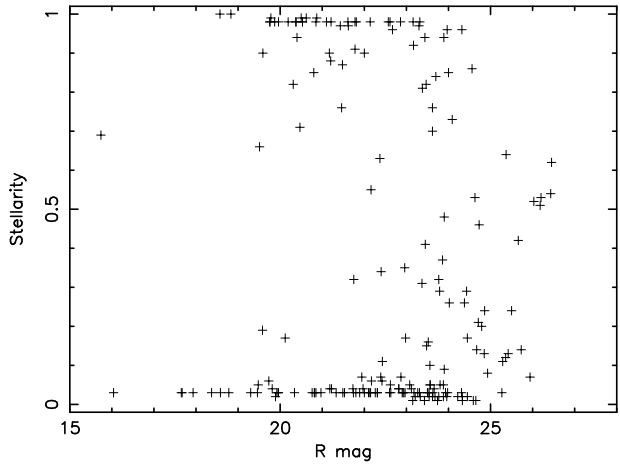}
\caption{Stellarity vs $R$-band magnitude for all {\em Chandra}
identifications.  }
\label{fig:id_stellarity}
\end{figure}

\begin{figure}
\psfig{figure=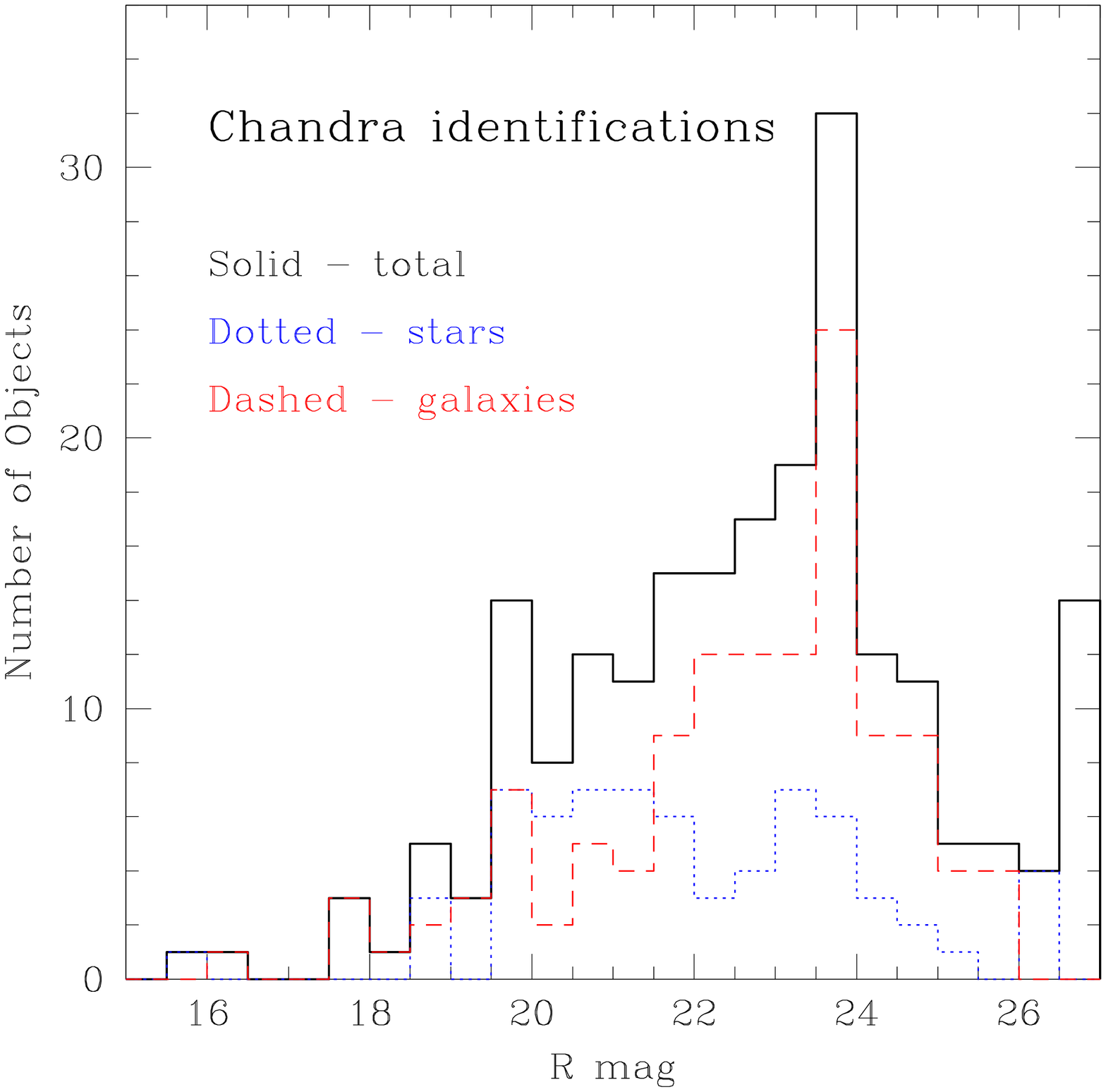,width=3.2in}
\caption{Magnitude distribution of {\em Chandra} identifications. 
`Stars' are defined, simplistically, to have stellarity $>0.5$ and 
galaxies have stellarity $\leq0.5$. Note that all sources with no
optical candidate have been placed in the faintest magnitude bin.
}
\label{fig:id_magoverview}
\end{figure}

We performed a simple morphological analysis of the identifications
using the stellarity parameter from the {\sc SExtractor} software.  In
Fig.~\ref{fig:id_stellarity} we show the distribution of stellarity
for the identifications listed in Table~\ref{tab:main}. The majority
of identifications are non-stellar. As the majority of X-ray sources
have a counterpart with $R<25$, {\sc SExtractor} is able to make a
reasonably confident judgment about their morphologies. Taking a crude
split between stars and galaxies at a stellarity level of 0.5, we
obtain Fig.~\ref{fig:id_magoverview}, which gives a summary of the
`stellar' (ie mostly QSO) and `galaxy' content of the survey.  We see
that the peak of the distribution for all morphological types is
around $R\sim23-24$.

At bright magnitudes ($R<22$) the galaxies appear visually to be
mainly ellipticals, although some definite spirals and lenticulars are
also seen.  At fainter magnitudes there are many low surface
brightness objects, which are probably disc galaxies, together with
some more compact objects which may be QSOs or compact galaxies.
These results are broadly consistent with the findings of other {\em
Chandra} surveys (eg Koekemoer \etal 2002) but we leave proper
morphological classification to a later paper.

We also note a number of cases where the optical postage stamps show
an apparent excess number of galaxies, eg sources 145 (perhaps
accounting for the large listed X-ray/optical offset), 162 and 192.
However as {\em Chandra} is not well suited to the detection of
extended low-surface brightness emission, we leave proper discussion
of the cluster content of our survey until later, when the {\em
XMM-Newton} data can be included.

\subsection{X-ray/Optical Flux Distributions}

\begin{figure*}
\psfig{figure=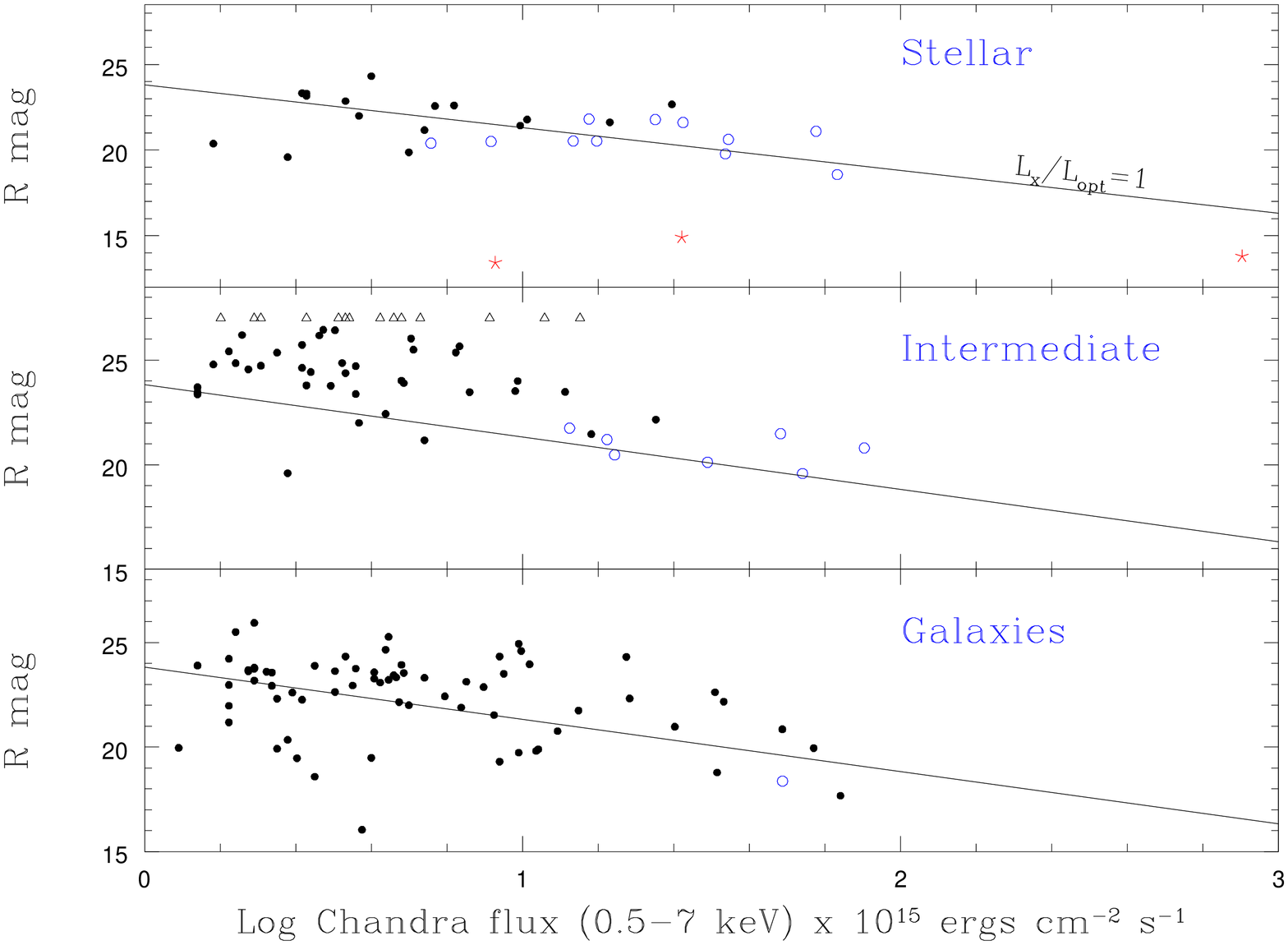,width=7in,angle=270}
\caption{$R$-band magnitude vs. {\em Chandra} flux
for the sources within $6'$ of a {\em Chandra} pointing axis.  The top
panel contains those counterparts which are most clearly classified as
stellar, ie with stellarity $>0.9$, the bottom panel contains those
counterparts which are most clearly classified as galaxies, ie with
stellarity $<0.1$, and the middle panel contains the remainder.
Sources with no optical counterpart (ie $R\gtsim27$) are shown in the
middle panel as open triangles. In all panels objects which have
previously been classified spectroscopically as QSOs, ie broad
emission line objects, are shown as open circles.  We note that one
nearby ($z=0.26$) broad line radio galaxy, where the host galaxy is
easily detectable, is thereby shown as a QSO in the `galaxy' panel.
In the top panel the three objects previously classified as galactic
stars are shown as asterisks. All other objects are filled circles.
In all panels a line representing an X-ray/optical ratio of 1, typical
of unabsorbed AGN, is shown.  }
\label{fig:magflux}
\end{figure*}

Previous authors (eg Giacconi \etal 2002) have shown that there is a
wide range of X-ray/optical ratios amongst the identifications in the
very deep \ch surveys. Here we present the X-ray/optical ratios for
the sources in our less deep survey (Fig.~\ref{fig:magflux}). We also
split our identifications into three morphological classes. In
Fig.~\ref{fig:magflux} we define `stellar' objects to have stellarity
$>0.9$, `galaxies' to have stellarity $<0.1$ and all other
stellarities are classified as `intermediate'. Using these rather
strict definitions, our `stellar' and `galaxy' identifications should
be unequivocal. In all panels we present objects which have previously
been classified spectroscopically as `QSOs' with open circles.  In all
panels we show the (unity) line of X-ray/optical ratio, $f_{x}/f_{opt}
= 1$, characteristic of unabsorbed AGN (eg Stocke \etal 1991). We
follow the most widely used definition of $f_{x}/f_{opt}$, as given by
Stocke \etal In Appendix B we discuss the conversion of the
$f_{x}/f_{opt}$ relationship between the optical and X-ray bands used
by Stocke and those used here. We also compare our definition of
$f_{x}/f_{opt}$ with the very similar, but not absolutely identical,
one of Giacconi \etal (2002).

A number of conclusions are obvious from examination of
Fig.~\ref{fig:magflux}.  Amongst the `stellar' identifications, the
galactic stars lie, as expected, well below the unity line.  With only
about 2 or 3 exceptions, all of the other stellar objects, including
the previously spectroscopically classified broad line objects, have
optical magnitudes within 2~mags of the unity line, in agreement with
the typical spread in $f_{x}/f_{opt}$ found by Stocke \etal (1991) for
broad line AGN.  Thus there are probably very few stars in our sample
and the QSOs are, in general, not greatly absorbed.  Incidentally, we
note a very good correspondence between morphological classification
as a stellar object via the stellarity indicator and spectral
classification as a QSO. See Tables~\ref{tab:main} and \ref{tab:rosat}
for more details.

The distribution of definite galaxies around the unity line is very
wide.  The objects below the line (mostly bright elliptical galaxies)
could be either low luminosity AGN, possibly including advective flow
AGN (di Matteo \& Fabian 1997) where the light from the host galaxy
dominates the optical light, starburst galaxies, or low luminosity BL
Lacs. However, typically pure starburst galaxies have $f_{x}/f_{opt} <
0.01$ and so would lie more than 5 magnitudes below the unity
line. Although some galaxies lie that far from the line, and so may be
starburst dominated, most do not and so, if they do contain a
starburst component, they must also contain another component of
higher $f_{x}/f_{opt}$, eg an AGN.

However most galaxies lie above the unity line, with an average
separation from the line of roughly twice that of the stellar objects.
A combination of absorption and moderately high redshift is required
to produce $f_{x}/f_{opt} > 1$. Absorption will preferentially remove
lower energy X-rays, but redshift will bring the less absorbed, higher
energies, into the observing band. However redshift brings the more
absorbed UV into the optical observing band. As the point-like active
nucleus is then not bright in the optical observing band, the optical
light will be dominated by the host galaxy and the object will have a
low stellarity. This conclusion is robust against chance
identifications as, if the optical identification listed in
Table~\ref{tab:main} is not correct then the correct identification
will be even fainter and so of higher $f_{x}/f_{opt}$.

Where redshifts are available for faint objects in other {\em Chandra}
surveys (eg Rosati \etal 2002; Crawford \etal 2002; Barger \etal
2001), values around unity are found, particularly for the type 2 AGN.
Although we do not yet have spectroscopic redshifts for the high
$f_{x}/f_{opt}$ galaxies in our survey, we can make crude redshift
estimates, assuming that the galaxies do host absorbed AGN and so the
host galaxies will typically be more luminous than $M^{*}$ (eg
Crawford \etal 2001). Assuming star formation histories and hence
evolutionary and K-corrections similar to those invoked to explain the
faint galaxy counts (Metcalfe \etal 2001), then for the faint galaxies
which lie mainly in the range $22<R<26$, redshifts in the range 0.5 to
2 (with slightly higher redshifts being allowable for late type
galaxies) are therefore generally appropriate unless the host is considerably
more luminous than $M^{*}$. For even fainter identifications, the
hosts must either be at even higher redshift, or be less luminous
than $M^{*}$.

The `intermediate' objects contain, at brighter X-ray and optical
fluxes, some spectroscopically confirmed QSOs which lie close to the
unity line. In the magnitude range $23-25$, where {\sc SExtractor }
can produce a reasonably reliable classification, almost all of the
intermediate objects have an optical FWHM $<1.5''$ compared to only
$\sim$half of the galaxy sample in the same magnitude range.  These
intermediate objects may therefore have contributions from both a host
galaxy and an unresolved AGN core to their optical light.  {\em
Chandra} counterparts with $R>25$ have unreliable morphological
classifications and are placed, by {\sc SExtractor}, within the
`intermediate' class.  These faint objects are mostly of high
$f_{x}/f_{opt}$ and so are more likely to be (absorbed) galaxies than
stellar objects.

Although our sources (of all stellarity combined) generally straddle
the unity line, more lie above it than below. Thus our distribution
differs, at first sight, from the deeper sample of Giacconi \etal
(2002) which has many sources of low $f_{x}/f_{opt}$ below an X-ray
flux of $1 \times 10^{-15}$ \ecs (0.5-10 keV), mostly associated with
nearby optically bright galaxies (cf Hornschemeier \etal 2001) which
we do not detect.  However above the limiting flux of our sample,
their distribution of $f_{x}/f_{opt}$ is similar to ours.

\section{Comparison with previous {\em ROSAT} Survey}
\label{sec:rosat}

\subsection{Accuracy of {\em ROSAT} Positions}
\label{sec:rospos}

We list, in Table~\ref{tab:rosat}, the {\em Chandra}, and {\em ROSAT},
positions of the {\em ROSAT} sources from \mch \etal (1998) which are
covered by the present {\em Chandra} observations.  For comparison,
the flux limit of the \ch survey, on axis, in the {\em ROSAT} band
($0.5-2$ keV), is $5.6 \times 10^{-16}$ \ecs (assuming $\alpha = 0.7$)
compared with the flux limit of the {\em ROSAT} survey of $2 \times
10^{-15}$ \ecs$\!$.  We searched for {\em Chandra} sources within a
pessimistically large radius of 20 arcsec of the {\em ROSAT} X-ray
position.  In total, 76 {\em ROSAT} sources were found to have a
corresponding \ch source.  {\em ROSAT} sources R9, R16, R29, R78, R84
and R100 were off the edge of any of the ACIS-I chips, sources R98,
R111 and R126 were on the very edge of a chip and sources R43, R105
and R123 were very far from the \ch axis ($9.4'$, $9.3'$ and $8.6'$
respectively).  These sources are therefore not listed in
Table~\ref{tab:rosat}.

Of the {\em ROSAT} sources not detected by \ch, but within the \ch
field of view (R34, R97, R108, R117, R129, R133 and R135), only R34,
R117 and R129 are within $6'$ of a \ch pointing and so might perhaps
be detectable. R34 was catalogued as a cluster and is clearly detected
as an extended source by {\em XMM-Newton} (the {\em XMM-Newton}
observations will be presented elsewhere; Mason \etalc, in
preparation; Page \etal 2003).  R117 is also easily detected by {\em
XMM-Newton}. There is a hint of an extension, although it is hard to
be certain, and the X-ray spectrum appears to be very soft.  R129 is
not detected by {\em XMM-Newton} but was detected by our (unpublished)
{\em ROSAT} HRI observations, indicating variability. (The HRI source
was $5''$ from the original {\em ROSAT} PSPC source but not on the
originally suggested identification.).  Thus {\em Chandra} detects
almost all of the {\em ROSAT} sources within its field of view. A
small amount of variability, together with reduced sensitivity far off
the {\em Chandra} axis, is sufficient to account for the rest. There
is no evidence that any of the {\em ROSAT} sources were false.

\begin{figure}
\psfig{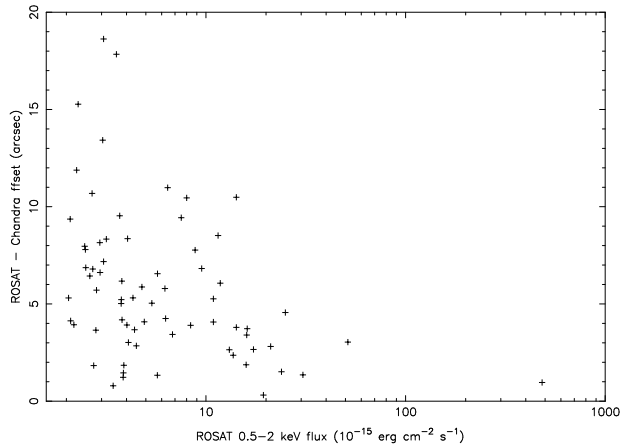}
\caption{Distribution of {\em ROSAT}/{\em Chandra} offsets as a
function of {\em ROSAT} flux.
}
\label{fig:rosatoffset}
\end{figure}

Within the $20''$ search radius around the {\em ROSAT} sources we
expect 8 random \ch coincidences, approximately half of which will lie
beyond $\sim15''$. Therefore the 3 sources which we note in
Fig.~\ref{fig:rosatoffset} beyond $\sim15''$ are probably random
coincidences and so should be removed before determining average {\em
ROSAT}--\ch offsets.  For 6 {\em ROSAT} sources (R5, R17, R47, R60,
R109 and R120) we detect 2 {\em Chandra} counterparts. With the
exception of R120, which is noted in \mch~\etal as being a potential
cluster candidate, one {\em Chandra} counterpart is considerably
closer to the {\em ROSAT} centroid and only the closer counterpart to
these six {\em ROSAT} sources is listed in Table \ref{tab:rosat}.

We note that R23, the one relatively bright {\em ROSAT} source where
{\em Chandra} finds a counterpart far off ($10''$), has 2 counterparts
in our HRI image. The second counterpart (not detected by \ch) is
associated with a NELG on the opposite side of the {\em ROSAT}
errorbox, a likelihood discussed in \mch~\etal Thus the {\em ROSAT}
source is confused and its centroid ended up partway between these 2
sources.

In Fig.~\ref{fig:rosatoffset} we plot the resultant {\em ROSAT}--{\em
Chandra} offsets as a function of {\em ROSAT} flux. Although it
makes little difference, we do not include sources which are
identified with clusters or groups as our present PSF-fitting source
searching, which is tuned to detect unresolved sources, is not well
suited to detecting or determining centroid positions for clusters.
We do, however, include sources at large {\em Chandra} off-axis angles
where {\em Chandra} positional errors can themselves be significant.
Similarly we include {\em ROSAT} sources at large {\em ROSAT} off-axis
angles.  Nonetheless the distribution of offsets, even though we 
retain sources at large {\em Chandra} and {\em ROSAT} off-axis
angles, and probably include some non-real coincidences at
large offsets, is almost exactly the same as we predicted in
\mch~\etal (1998, fig.~4); eg the mean offset at the lowest {\em
ROSAT} fluxes (2-3 $\times 10^{-15}$ ergs cm$^{-2}$ s$^{-1}$) is 7.2
arcsec (which is reduced to 6.7 arcsec if we omit the potentially
spurious source with a {\em ROSAT/Chandra} offset $> 15''$), with a
90 per cent value of $\sim10$ arcsec. We therefore confirm the
accuracy of the positions in our earlier {\em ROSAT} survey.

\noindent 
\begin{table*} 
\caption{Catalogue of \ROSAT sources detected by \Chandra, showing their 
respective coordinates, the offset between the X-ray positions, and the offset between the \Chandra source and the \ROSAT optical counterpart (in the Subaru coordinate frame).  The offaxis angle of each X-ray source in both the \Chandra and \XMM/\ROSAT observations are listed, together with the classification, redshift and confidence parameter (Rflag), taken from M$^{\rm c}$Hardy \etal (1998).  The Conf parameter shows whether the optical counterpart to the \ROSAT source has been confirmed or not: Y denotes that the optical counterpart is $<1''$ from the \Chandra source when OffC $< 6'$ (extended to $<1.6''$ for $6'<$ OffC $<8'$); N denotes that the optical counterpart is $>3''$ from the
\Chandra source; and ? denotes all intermediate cases.} 
\begin{center} 
\begin{tabular}{rccrccrrrrlccc} 
ROS & \multicolumn{2}{c}{\ROSAT} & Ch & \multicolumn{2}{c}{\em Chandra} & $\delta^{\rm Tot}_{R-C}$ & $\delta^{\rm Tot}_{O-C}$ & OffC & OffX & Class & $z$ & Rflag & Conf \\ 
Num & \multicolumn{2}{c}{RA (J2000) Dec} & Num & \multicolumn{2}{c}{RA (J2000) Dec} & ($''$) & ($''$) & ($'$) & ($'$) & & & & \\ 
  1 & 13:34:51.47 & 37:46:19.9 &   1 & 13:34:51.43 & 37:46:19.1 &  1.0 &  1.0 &  5.5 &  8.9 &  MSTAR & 0.00 &   * &   Y \\ 
  2 & 13:34:41.66 & 38:00:09.1 &  10 & 13:34:41.83 & 38:00:11.4 &  3.0 &  0.2 &  4.0 &  5.5 &    QSO & 0.26 &   * &   Y \\ 
  3 & 13:33:42.32 & 38:03:34.6 &   2 & 13:33:42.28 & 38:03:35.9 &  1.4 &  1.4 &  7.9 & 13.9 &    QSO & 1.07 &   * &   Y \\ 
  5 & 13:34:37.72 & 37:56:05.3 &   8 & 13:34:38.08 & 37:56:03.9 &  4.6 &  0.2 &  5.4 &  1.4 &    CLU & 0.57 & (*) &   Y \\ 
  7 & 13:34:17.43 & 37:57:22.1 &   5 & 13:34:17.55 & 37:57:22.6 &  1.5 &  0.3 &  1.7 &  4.7 &    QSO & 1.14 &   * &   Y \\ 
 10 & 13:34:10.64 & 37:59:55.9 &   9 & 13:34:10.63 & 37:59:56.1 &  0.3 &  0.3 &  1.3 &  7.3 &    QSO & 0.38 &   * &   Y \\ 
 11 & 13:33:31.98 & 37:46:39.3 &  14 & 13:33:32.03 & 37:46:42.1 &  2.8 &  0.9 & 10.0 & 15.1 &    QSO & 0.83 &   * &   Y \\ 
 13 & 13:33:58.58 & 37:59:35.8 &   7 & 13:33:58.55 & 37:59:38.5 &  2.7 &  0.4 &  3.2 &  9.0 &    QSO & 1.61 &   * &   Y \\ 
 15 & 13:34:42.50 & 37:59:16.3 &  18 & 13:34:42.77 & 37:59:15.0 &  3.4 &  0.2 &  3.9 &  4.7 &    QSO & 1.14 &   * &   Y \\ 
 17 & 13:34:01.03 & 37:54:01.2 &  23 & 13:34:01.04 & 37:54:05.0 &  3.7 &  0.1 &  5.5 &  7.1 &    QSO & 1.64 & (*) &   Y \\ 
 18 & 13:35:44.55 & 37:51:39.8 &  12 & 13:35:44.70 & 37:51:40.4 &  1.9 &  0.9 &  7.4 & 13.7 &    QSO & 1.62 &   * &   Y \\ 
 20 & 13:35:30.06 & 37:57:51.1 &  30 & 13:35:30.35 & 37:57:49.5 &  3.8 &  0.8 &  6.0 & 11.0 &    QSO & 1.39 &   * &   Y \\ 
 21 & 13:34:31.16 & 37:48:31.4 &   3 & 13:34:31.35 & 37:48:31.5 &  2.4 &  0.2 &  2.4 &  6.3 &    QSO & 1.36 &   * &   Y \\ 
 23 & 13:33:44.86 & 37:57:60.0 &  29 & 13:33:44.23 & 37:57:52.5 & 10.5 &  0.8 &  6.1 & 10.9 &    QSO & 0.97 & (*) &   Y \\ 
 24 & 13:35:35.31 & 37:57:45.1 &  15 & 13:35:35.53 & 37:57:45.9 &  2.6 &  0.7 &  7.0 & 11.9 &    QSO & 1.63 &   * &   Y \\ 
 27 & 13:34:09.05 & 38:03:44.5 &  25 & 13:34:08.76 & 38:03:49.4 &  6.1 &  0.4 &  5.0 & 10.7 &  GSTAR & ---  &   * &   Y \\ 
 30 & 13:34:51.74 & 37:57:45.8 &  22 & 13:34:52.17 & 37:57:44.7 &  5.3 &  0.1 &  3.0 &  4.2 &    QSO & 1.89 &   * &   Y \\ 
 31 & 13:33:55.70 & 37:52:54.7 &  17 & 13:33:55.81 & 37:52:58.5 &  4.1 &  0.3 &  5.9 &  8.3 &    QSO & 2.14 &   * &   Y \\ 
 32 & 13:35:24.75 & 38:05:36.3 &  16 & 13:35:25.45 & 38:05:34.3 &  8.5 &  1.5 &  7.2 & 14.4 &   NELG & 0.07 &   * &   Y \\ 
 36 & 13:34:38.20 & 38:06:20.7 &  33 & 13:34:38.51 & 38:06:26.5 &  6.8 &  0.7 &  7.9 & 11.7 & G/NELG & 0.23 &   * &   Y \\ 
 37 & 13:34:23.96 & 37:46:16.1 &  38 & 13:34:24.61 & 37:46:15.3 &  7.8 &  0.5 &  3.4 &  8.8 &    QSO & 1.57 &   * &   Y \\ 
 42 & 13:35:02.90 & 37:50:00.5 &  21 & 13:35:02.85 & 37:49:56.6 &  3.9 & 14.8 &  1.2 &  7.0 &   NELG & 0.37 &   * &   N \\ 
 47 & 13:35:05.42 & 37:49:54.3 &  20 & 13:35:06.21 & 37:49:52.9 &  9.4 &  0.5 &  0.9 &  7.5 &   NELG & 0.36 &   * &   Y \\ 
 48 & 13:35:29.05 & 38:04:26.5 &   6 & 13:35:29.77 & 38:04:32.6 & 10.5 &  1.3 &  7.1 & 14.3 &    QSO & 0.69 &   * &   Y \\ 
 49 & 13:34:46.81 & 37:47:52.2 &  11 & 13:34:46.94 & 37:47:48.5 &  4.0 &  0.2 &  5.0 &  7.2 &   CLU? & 0.71 &  -- &   Y \\ 
 51 & 13:34:00.15 & 37:49:10.2 &   4 & 13:33:59.91 & 37:49:12.1 &  3.4 & 24.3 &  4.1 &  9.2 &   NELG & 0.06 & (*) &   N \\ 
% 54 & 13:33:22.64 & 37:57:47.8 &  44 & 13:33:24.32 & 37:57:45.4 & 20.0 & --- & 10.0 & 14.6 &      ? & ---  &  -- & -- \\ 
 55 & 13:34:47.17 & 37:59:46.5 &  13 & 13:34:47.37 & 37:59:50.1 &  4.3 &  0.5 &  2.9 &  5.5 &    QSO & 1.18 &   * &   Y \\ 
 56 & 13:34:44.87 & 37:57:24.2 &  53 & 13:34:45.34 & 37:57:22.7 &  5.8 &  0.4 &  4.2 &  3.1 &    QSO & 1.89 &   * &   Y \\ 
 57 & 13:33:35.51 & 37:54:11.3 &  62 & 13:33:35.61 & 37:54:00.4 & 11.0 & 13.0 &  9.1 & 12.1 &    QSO & 1.52 &   * &   N \\ 
 58 & 13:34:34.00 & 37:57:03.2 & 203 & 13:34:35.00 & 37:56:50.0 & 17.6 & 15.2 &  4.5 &  2.1 &    CLU & 0.31 &   * &   N \\ 
 60 & 13:34:08.86 & 37:57:05.7 &  28 & 13:34:08.82 & 37:57:06.9 &  1.3 &  0.2 &  2.1 &  6.0 &  NELG? & 0.58 &     &   Y \\ 
 61 & 13:33:34.87 & 37:49:16.9 &  49 & 13:33:34.49 & 37:49:12.1 &  6.6 &  0.3 &  9.1 & 13.5 &    QSO & 3.43 &   * &   Y \\ 
 62 & 13:35:09.32 & 37:48:21.7 & 100 & 13:35:09.75 & 37:48:20.2 &  5.3 & 14.3 &  2.4 &  9.1 & GP/GAL & 0.25 &   * &   N \\ 
 63 & 13:34:22.27 & 38:06:16.1 &  41 & 13:34:22.17 & 38:06:21.0 &  5.0 &  1.9 &  7.6 & 12.0 &    QSO & 2.59 &   * &   ? \\ 
 64 & 13:34:14.48 & 37:51:28.2 &  81 & 13:34:14.71 & 37:51:31.3 &  4.1 &  0.2 &  2.2 &  5.5 &  GSTAR & ---  &   * &   Y \\ 
 67 & 13:34:59.90 & 37:56:28.1 &  70 & 13:35:00.20 & 37:56:32.8 &  5.9 &  0.4 &  3.5 &  4.9 &   NELG & 0.56 &   * &   Y \\ 
 72 & 13:35:15.29 & 37:58:35.6 &  83 & 13:35:15.24 & 37:58:38.4 &  2.8 &  0.4 &  3.0 &  8.5 &    QSO & 2.81 &   * &   Y \\ 
 73 & 13:35:16.99 & 37:54:18.9 &  86 & 13:35:17.33 & 37:54:15.5 &  5.3 &  0.6 &  4.0 &  8.0 & G/QSO? & ---  &     &   Y \\ 
 74 & 13:34:07.63 & 38:06:20.2 &  60 & 13:34:08.20 & 38:06:28.5 & 10.6 & 10.2 &  7.7 & 13.0 &    CLU & 0.38 &   * &   N \\ 
 75 & 13:35:17.36 & 38:02:45.7 &  32 & 13:35:17.64 & 38:02:47.3 &  3.7 &  0.3 &  4.1 & 11.3 &    QSO & 1.38 &   * &   Y \\ 
 77 & 13:35:32.63 & 37:45:49.0 &  98 & 13:35:31.98 & 37:45:36.0 & 15.1 &  1.0 &  7.0 & 14.2 &    GRP & 0.31 & (*) &   Y \\ 
 80 & 13:34:11.20 & 37:47:54.6 &  66 & 13:34:11.28 & 37:47:57.5 &  3.0 &  0.4 &  2.4 &  8.5 &   GAL? & 0.33 &     &   Y \\ 
 82 & 13:35:15.82 & 37:52:41.1 &  19 & 13:35:15.92 & 37:52:40.7 &  1.2 &  0.1 &  2.5 &  8.0 &   QSO? & ---  &     &   Y \\ 
 85 & 13:34:08.40 & 37:54:42.5 &  58 & 13:34:08.54 & 37:54:41.8 &  1.8 &  4.8 &  4.4 &  5.6 &   NELG & 0.30 &   * &   N \\ 
 87 & 13:33:43.02 & 37:45:18.9 &  46 & 13:33:43.28 & 37:45:11.1 &  8.4 & --- &  8.6 & 14.3 & BLANK? & ---  &  -- & -- \\ 
 88 & 13:33:37.27 & 37:47:59.7 &  73 & 13:33:37.54 & 37:47:57.3 &  3.9 & --- &  8.6 & 13.6 &      ? & ---  &  -- & -- \\ 
 90 & 13:35:12.58 & 37:44:18.2 &  34 & 13:35:12.69 & 37:44:18.7 &  1.5 &  1.1 &  6.5 & 12.6 &   QSO? & ---  &  -- &   Y \\ 
 91 & 13:34:58.10 & 38:04:26.4 & 103 & 13:34:58.38 & 38:04:30.4 &  5.2 &  0.5 &  4.5 & 10.6 &    QSO & 2.01 &   * &   Y \\ 
 92 & 13:34:01.00 & 37:46:52.5 &  48 & 13:34:01.16 & 37:46:47.9 &  5.0 &  0.3 &  4.7 & 10.6 &    QSO & 1.59 &   * &   Y \\ 
 93 & 13:33:53.16 & 38:02:00.1 &  72 & 13:33:53.53 & 38:02:04.5 &  6.2 &  9.2 &  5.2 & 11.3 &   NELG & 0.60 &   * &   N \\ 
 94 & 13:33:46.42 & 38:00:26.1 &  78 & 13:33:46.56 & 38:00:22.2 &  4.2 &  7.7 &  5.7 & 11.4 &   NELG & 0.06 &   * &   N \\ 
 96 & 13:34:58.47 & 37:50:23.2 & 160 & 13:34:58.83 & 37:50:17.7 &  7.0 & 21.3 &  1.8 &  6.2 &    CLU & 0.38 & (*) &   N \\ 
 99 & 13:35:03.70 & 37:44:27.3 &  61 & 13:35:03.71 & 37:44:09.4 & 17.8 & --- &  6.6 & 11.8 &  BLANK & ---  &  -- & -- \\ 
101 & 13:34:02.50 & 37:51:29.3 &  43 & 13:34:02.56 & 37:51:29.5 &  0.8 &  0.3 &  4.0 &  7.5 &    QSO & 1.35 &   * &   Y \\ 
103 & 13:33:30.87 & 37:48:06.1 &  56 & 13:33:30.72 & 37:48:15.5 &  9.5 &  7.7 &  9.9 & 14.6 &   NELG & 0.20 &   * &   N \\ 
104 & 13:34:31.63 & 37:49:58.6 &  55 & 13:34:31.24 & 37:49:53.1 &  7.2 &  0.3 &  2.1 &  5.0 &    QSO & 1.49 &   * &   Y \\ 
107 & 13:33:51.20 & 37:49:45.7 & 128 & 13:33:50.50 & 37:49:46.9 &  8.3 & --- &  5.9 & 10.4 &      ? & ---  &  -- & -- \\ 
109 & 13:33:39.87 & 37:52:26.6 & 143 & 13:33:40.73 & 37:52:42.1 & 18.6 &  2.1 &  8.4 & 11.3 &   QSO? & 2.12 &  -- &   ? \\ 
\end{tabular} 
\label{tab:rosat} 
\end{center} 
\end{table*} 
\addtocounter{table}{-1} 
\begin{table*} 
\caption{(cont.) Catalogue of \ROSAT sources detected by \Chandra, showing their respective coordinates, the offset between the X-ray positions, and the offset between the \Chandra source and the \ROSAT optical counterpart (in the Subaru coordinate frame).  The offaxis angle of each X-ray source in both the \Chandra and \XMM/\ROSAT observations are listed, together with the classification, redshift and confidence parameter (Rflag), taken from M$^{\rm c}$Hardy \etal (1998).  The Conf parameter shows whether the optical counterpart to the \ROSAT source has been confirmed or not: Y denotes that the optical counterpart is $<1''$ from the \Chandra source when OffC $< 6'$ (extended to $<1.6''$ for $6'<$ OffC $<8'$); N denotes that the optical counterpart is $>3''$ from the \Chandra source; and ? denotes all intermediate cases.} 
\begin{center} 
\begin{tabular}{rccrccrrrrlccc} 
ROS & \multicolumn{2}{c}{\ROSAT} & Ch & \multicolumn{2}{c}{\em Chandra} & $\delta^{\rm Tot}_{R-C}$ & $\delta^{\rm Tot}_{O-C}$ & OffC & OffX & Class & $z$ & Rflag & Conf \\ 
Num & \multicolumn{2}{c}{RA (J2000) Dec} & Num & \multicolumn{2}{c}{RA (J2000) Dec} & ($''$) & ($''$) & ($'$) & ($'$) & & & & \\ 
110 & 13:35:12.13 & 38:02:42.5 &  42 & 13:35:12.54 & 38:02:46.9 &  6.6 &  0.8 &  3.4 & 10.7 &    QSO & 1.85 &   * &   Y \\ 
112 & 13:34:35.85 & 37:54:22.7 &  94 & 13:34:35.86 & 37:54:19.0 &  3.7 & --- &  5.6 &  0.5 & BLANK? & ---  &  -- & -- \\ 
113 & 13:34:27.64 & 37:41:35.1 & 104 & 13:34:28.58 & 37:41:27.7 & 13.4 & --- &  8.3 & 13.4 &      ? & ---  &  -- & -- \\ 
115 & 13:35:22.70 & 37:49:16.6 &  36 & 13:35:22.83 & 37:49:11.1 &  5.7 & 21.5 &  3.4 & 10.6 &  MSTAR & ---  &   * &   N \\ 
116 & 13:34:54.55 & 38:08:01.5 & 109 & 13:34:54.02 & 38:07:56.3 &  8.2 & --- &  8.0 & 13.6 &      ? & ---  &  -- & -- \\ 
118 & 13:34:23.88 & 37:42:57.1 & 110 & 13:34:24.02 & 37:42:57.8 &  1.8 &  0.5 &  6.7 & 12.1 &    QSO & 1.00 &   * &   Y \\ 
119 & 13:34:14.56 & 37:52:26.1 & 119 & 13:34:14.27 & 37:52:31.6 &  6.4 & 15.3 &  3.2 &  5.0 & KSTAR? & ---  &  -- &   N \\ 
120 & 13:35:19.35 & 37:43:06.3 &  77 & 13:35:19.35 & 37:43:18.1 & 11.8 & --- &  7.8 & 14.2 &      ? & ---  &  -- & -- \\ 
120 & 13:35:19.35 & 37:43:06.3 &  96 & 13:35:19.24 & 37:42:59.6 &  6.8 & --- &  8.1 & 14.4 &      ? & ---  &  -- & -- \\ 
121 & 13:35:17.91 & 37:55:32.6 & 155 & 13:35:18.49 & 37:55:33.2 &  6.9 &  3.2 &  5.3 &  8.2 &   NELG & 0.31 &   * &   N \\ 
122 & 13:34:13.49 & 38:07:12.9 &  88 & 13:34:13.77 & 38:07:23.0 & 10.7 & --- &  8.5 & 13.4 &      ? & ---  &  -- & -- \\ 
124 & 13:34:21.33 & 38:04:46.0 & 133 & 13:34:21.89 & 38:04:50.3 &  8.0 & --- &  6.1 & 10.5 &      ? & ---  &  -- & -- \\ 
125 & 13:33:37.15 & 37:56:37.4 & 105 & 13:33:37.05 & 37:56:29.7 &  7.8 &  1.4 &  7.8 & 11.9 &    QSO & 1.68 &   * &   Y \\ 
127 & 13:34:57.11 & 37:49:40.1 & 112 & 13:34:57.28 & 37:49:43.5 &  3.9 & 11.5 &  2.3 &  6.4 &  NELG? & 0.25 & (*) &   N \\ 
128 & 13:34:49.84 & 38:06:54.8 &  92 & 13:34:50.74 & 38:07:05.7 & 15.3 & 22.3 &  7.4 & 12.7 &   GAL? & 0.26 &  -- &   N \\ 
130 & 13:33:43.40 & 37:50:32.0 &  79 & 13:33:42.48 & 37:50:27.4 & 11.9 & --- &  7.5 & 11.6 &  BLANK & ---  &  -- & -- \\ 
131 & 13:35:19.28 & 37:58:24.8 & 101 & 13:35:20.07 & 37:58:24.1 &  9.4 & 20.3 &  3.9 &  9.2 &  NELG? & 0.58 & (*) &   N \\ 
132 & 13:34:46.32 & 37:58:44.4 & 167 & 13:34:46.61 & 37:58:40.3 &  5.3 &  0.7 &  3.3 &  4.4 &   NELG & 0.22 &   * &   Y \\ 
134 & 13:34:01.06 & 38:01:28.5 &  84 & 13:34:00.87 & 38:01:25.0 &  4.1 & 25.1 &  3.7 &  9.8 &   NELG & 0.25 & (*) &   N \\ 
\end{tabular} 
\end{center} 
\end{table*}

\subsection{Accuracy of {\em ROSAT} Identifications}

In Table~\ref{tab:rosat} we list the offset between the {\em Chandra}
position and the position of the optical counterpart as found on the
Subaru image (or APM plate for bright stars).  The optical coordinate
frame of our earlier CFHT CCD imaging as listed in \mch~\etal (1998)
was based on FK5 stars scanned on POSS I survey plates, and the
current Subaru coordinate frame is based on FK5 stars scanned on POSS
II survey plates. Nonetheless the positions of unsaturated stellar
objects on the CFHT and Subaru images agree to better than $\sim 0.3$
arcsec.  The {\sc SExtractor} analysis on the Subaru image, however,
produces a more reliable centroid position for extended sources than
did our earlier analysis.  Taking into account variations due to
off-axis {\em Chandra} angle, we can safely confirm any {\em ROSAT}
identifications where the ({\em Chandra} - \ros optical position)
offset is $<1.3''$. The majority of the identifications are therefore
either unambiguously correct, or incorrect.  Only two sources have
uncertain optical counterparts: R63 (C41) and R109 (C143).  R63 is
optically identified with a $z=2.6$ QSO, but is $7.6'$ offaxis in the
\ch observation, and therefore the positional accuracy may be worse in
this case.  R109 is listed in \mch~\etal as being in a very confused
region in the optical, and therefore X-ray source confusion may be a
problem here, but is observed at $8.4'$ from the \ch axis which may
also contribute to the larger {\em ROSAT}--\ch offset.

There has been considerable discussion regarding the reality of the
NELGs as identifications in \mch~\etal and so we consider them further
here.  A variety of confidence classes were assigned to
identifications in \mch~\etal so let us consider the highest
confidence class, designated by `*'.  Although most of the
identifications are either unambiguously correct or incorrect, there
is some uncertainty regarding R43, R85 and R117.  In the case of R85
there is moderately strong radio emission (Seymour \etalc, in
preparation) from the optical NELG. However the {\em Chandra} position
is too far ($4.8''$) from the optical galaxy for it to be considered
the major source of the X-rays, although it could be a lesser
contributor, and so for these purposes we consider it an incorrect
identification.  In very preliminary analysis of the {\em XMM-Newton}
observations (Mason \etalc, in preparation) very soft spectrum X-ray
emission comes from the optical position of R117 and, in both the {\em
Chandra} and {\em XMM-Newton} images, no other source of X-ray
emission is detectable in the {\em ROSAT} errorbox.  R117 is a large
($\sim10$arcsec) galaxy (see Gunn \etal 2001), with extended radio
emission, and the probability is that {\em XMM-Newton} is detecting
extended X-ray emission which our current \ch source detection
algorithm is not tuned to detect, in this faint source. We therefore
consider it a correct identification with starburst emission being a
major contributor to its X-ray emission, consistent with its very low
$f_{x}/f_{opt}$ ratio. R43 is off the {\em Chandra} field and towards
the edge of the {\em XMM-Newton} field, so sensitivity is low. There
is a very faint source in the {\em XMM-Newton} image, close to the
proposed optical counterpart to R43, but its position cannot be
determined to better than 4 arcsec. However R43 coincides with by far
the brightest radio source in the {\em ROSAT} field. We therefore
class R43 as probably a correct identification.

Thus of the higher confidence NELGS, 7/13 are confirmed although the
lower confidence NELGs are generally not confirmed.  The incorrect
\ros identifications, of all types, mostly lie at large distances from
the \ros centroid and so do not have the highest confidence
classification. The number of chance NELG misidentifications is a
little larger than initially estimated ($\sim$ half as opposed to
$\sim$ one third of the sample) but not outside of reasonable
limits. The NELGS, as a class, are certainly confirmed as being
significant emitters of soft X-rays although, physically, they are
probably a mixture of AGN and starburst, even within the same galaxy,
as we stated in \mch \etal (1998).

We also note that not all QSOs were confirmed as the correct
identification by {\em Chandra}, eg R57. Although variability is the
more likely physical explanation, we should perhaps not automatically
assume that every QSO that is found in an X-ray errorbox is the only
source of the emission.

\subsection{Differences between the \ch and {\em ROSAT} sources}

We have shown above (Section~\ref{sec:rospos}) that \ch detects all of
the {\em ROSAT} sources that one might reasonably have expected it to
detect.  The {\em ROSAT} survey limit of $2\times 10^{-15}$ \ecs
($0.5-2$ keV) corresponds to a flux of $4.7 \times 10^{-15}$ \ecs in
the {\em Chandra} $0.5-7$ keV band, assuming an average source energy
spectral index of $\alpha = 0.7$.  In our {\em Chandra} sample, the
brightest 126 sources lie above this flux limit, of which 74 are
detected by {\em ROSAT}.  Approximately two thirds of the `stellar'
{\em Chandra} sources ($S>0.9$) are detected by {\em ROSAT} compared
to only about a half of those sources with optically resolved
($S<0.1$) counterparts.  \mch \etal (1998) found that only $\sim$20\%
of the \ros sources showed detectable flux variability between the
\ros AO1 and AO4 observations. Thus variability cannot be the major
reason for the difference. The simplest reason for the difference is
absorption in some of the \ch sources, which would not be surprising
in view of the hard spectrum of the XRB. To crudely quantify the
absorption needed, an absorbing column of $N_{H} \sim 2\times 10^{22}$
atoms cm$^{-2}$ is equivalent to an exponential cut-off energy of 2
keV (the energy usually taken as the upper bound of the standard \ros
PSPC band).  For standard gas to dust ratios, that column is also
equivalent to an optical obscuration of $A_{R} \sim 9$mag, which is
sufficient to obscure optical emission from the AGN. Thus obscuration
is entirely consistent with the fact that \ros preferentially detects
objects where the point-like optical emission from the nucleus is
visible.  Indeed Piconcelli \etal (2002) and Mainieri \etal (2002)
demonstrate the presence of absorption at approximately the required
level in deep \xmm surveys.  Page \etal (2003) also confirm absorption
in some of our present \ch sources.  A full treatment of the spectral
and variability properties of the present \ch sample, based on the
\xmm data, will be presented in future papers.

\section{Conclusions}

We detect 214 \ch sources above a likelihood value of 25
(approximately $5 \sigma$) above a flux of $\sim 1.3 \times 10^{-15}$
\ecs ($0.5-7$ keV).  The fluxes are typical of the sources which provide
the bulk of the $0.5-7$ keV XRB.  The positional accuracy of our
survey has been investigated thoroughly, and quantified.  For the 151
sources which lie within $6'$ of a \ch pointing axis, the positions
are good to $\sim 0.7''$ and we expect very few ($\sim2$) chance
coincidences brighter than $R=24$. At larger off-axis angles the
positional accuracy decreases, although the majority of the remaining
sources are still expected to be correctly identified.  The optical
identification content of our sample is similar to that of the very
deep {\em Chandra} surveys (eg Giacconi \etal 2002) to our X-ray flux
limit. We do not detect, in large numbers, the very low luminosity
X-ray galaxies which are detected at lower X-ray fluxes in the deepest
surveys. However, owing to the larger area of our survey compared to
the typical very deep survey, we provide relatively large numbers of
sources with fluxes around the knee in the medium energy source
counts.  We briefly summarise our optical content here.

The very large majority of the X-ray sources have an optical
counterpart, with the distribution peaking at $23<R<24$. However a
small number have no counterpart to $R=27$.  Elliptical galaxies
dominate the bright ($R\leq21$) identifications (cf Mushotzky \etal
2000) but both high and low surface brightness objects are found at
fainter magnitudes. A useful summary of our optical identification
content is provided by Fig.~\ref{fig:magflux} which combines
X-ray/optical flux ratios with the simple `Stellarity' optical
morphological parameter.  For any given X-ray flux, and particularly
at fainter fluxes, there is a very large spread ($\sim7$ mags) in the
optical magnitudes of the `galaxy' counterparts (less so for the
point-like identifications), indicating a wide range of emission
mechanisms.  For the faintest optical galaxies, absorption together
with a moderate/high redshift is required to produce the observed
X-ray/optical ratios. A redshift range of, very roughly, $0.5-2$, is
consistent with the observed galaxy magnitudes.  and the estimated
luminosities are only moderate ($\ltsim 10^{43}$ ergs s$^{-1}$).
Although most of the \ch counterparts are classed as galaxies or
intermediate, a non-negligible number of definitely `stellar'
counterparts are found.  Galactic stars clearly show up as having
lower X-ray/optical ratios than the other stellar objects which
cluster around the line of $f_{x}/f_{opt} =1$, indicating that they
are probably mostly relatively unabsorbed QSOs.

Excluding the confirmed galactic stars and bright galaxies of low
$f_{x}/f_{opt}$, there are more galaxies than stellar objects. If, as
argued earlier, the galaxies are AGN in which the optical source is
obscured, these results are consistent with the greater numbers of
type 2, compared to type 1, Seyfert galaxies found in optical samples
(Maiolino \& Rieke 1995) and with spectral synthesis models of the
X-ray background (eg Comastri \etal 1995; Gilli \etal 2001).

The positions of the X-ray sources from our {\em ROSAT} survey were
shown to be in excellent agreement with the predictions based on the
simulations in \mch \etal (1998). \ch detects all of the {\em ROSAT}
sources that one might reasonably have expected it to detect
(Section~\ref{sec:rospos}). There is no evidence that any of the {\em
ROSAT} sources were false. Most of the {\em ROSAT} identifications are
confirmed.  In particular 7 of the 13 highest confidence
identification NELGs are confirmed. The X-ray emission in the more
luminous NELGs is probably dominated by AGN but the lowest luminosity
NELG is starburst dominated.

If, conversely, we ask how many of the \ch sources we would have
expected {\em ROSAT} to detect then, assuming $\alpha=0.7$ and no
absorption, we expect 126 compared to 74 actually detected. \ch
sources with `stellar' optical counterparts were preferentially
detected by \ros. The different numbers, and the preferential detection of
sources with `stellar' counterparts is probably due to intrinsic
absorption in many of the \ch sources and is not particularly
surprising in light of the `hard' spectrum of the medium energy XRB.
Page \etal (2003) present intial \xmm spectra of our \ch sources which
confirms the existence of substantial absorption. More detailed
spectra will be presented elsewhere.

The sources presented here, with fluxes near the `knee' of the medium
energy X-ray source counts, constitute an excellent sample on which to
study the astrophysics of the contributors to the XRB and the
relationship of those contributors to the dominant populations in
other wavebands. Initial papers are presented by Page \etal (2003) and
Gunn \etal (2003) and other are in preparation. These papers rely on the
accurate optical identification of the X-ray sources, which is the
main aim of the present paper.

\section*{Acknowledgments}

We thank the staff of the {\em Chandra} Observatory for carrying out
the observations so well, including ensuring identical roll angles for
all observations, and for providing final data products only 4 days
after the observations. We also thank the Subaru observatory for the
$R$-band SuprimeCam image.  We thank Mike Irwin for rapidly obtaining,
and scanning, the second epoch Palomar plates for us. We thank Tom
Shanks for providing a tabular version of K and evolutionary
corrections.  This work was supported by grants to a number of authors
from the UK Particle Physics and Astronomy Research Council (PPARC)
including PPA/G/S/1999/00102 and also by NASA grant G01-2098X.  IMcH
also acknowledges the support of a PPARC Senior Research Fellowship.

\appendix

\section{Notes on sources with more than one possible counterpart}

\begin{table*} 
\caption{Chandra sources with alternative optical counterparts.} 
\begin{center} 
\begin{tabular}{rrrrrrrrrrrcc} 
 Num & Flux & \multicolumn{1}{c}{$\chi^{2}$} & \multicolumn{2}{c}{Chandra} & \multicolumn{1}{c}{$\delta_{O-X}^{\rm RA}$} & \multicolumn{1}{c}{$\delta_{O-X}^{\rm Dec}$} & \multicolumn{1}{c}{$\delta_{O-X}^{\rm Tot}$}  & \multicolumn{1}{c}{R} & \multicolumn{1}{c}{OffC} & \multicolumn{1}{c}{OffX} & FWHM & Stellar \\ 
&&& \multicolumn{2}{c}{RA (J2000) Dec} & \multicolumn{1}{c}{($''$)} & \multicolumn{1}{c}{($''$)} & \multicolumn{1}{c}{($''$)} & (mag) & \multicolumn{1}{c}{($'$)} & \multicolumn{1}{c}{($'$)} & ($''$) & \\ 
&&&&&&&&&&&& \\ 
  47 &   2.07 &   152.3 & 13:33:32.09 & 37:58:15.86 & -0.68 &  1.35 & 1.51 &  25.3 &   8.40 & 13.27 &   1.59 & 0.36 \\  
  56 &   1.81 &    90.7 & 13:33:30.72 & 37:48:15.50 & -0.04 & -0.79 & 0.79 &  24.8 &   9.91 & 14.61 &   1.78 & 0.01 \\  
  57 &   1.79 &    70.5 & 13:33:20.30 & 37:57:44.85 & -1.95 & -0.29 & 1.97 &  23.4 &  10.77 & 15.41 &   2.12 & 0.03 \\  
  85 &   1.09 &   150.3 & 13:34:13.07 & 37:58:30.96 & -0.17 & -1.83 & 1.84 &  23.7 &   0.51 &  6.05 &   2.63 & 0.02 \\  
  95 &   0.91 &   130.6 & 13:34:37.20 & 37:54:36.78 & -1.42 &  1.24 & 1.89 &  27.2 &   6.01 &  0.13 &   1.16 & 0.50 \\  
 102 &   0.80 &    54.0 & 13:35:43.14 & 37:53:08.69 & -0.43 & -1.06 & 1.15 &  23.6 &   7.39 & 13.15 &   1.23 & 0.30 \\  
 115 &   0.70 &    60.8 & 13:35:36.41 & 37:51:11.58 &  1.92 &  0.42 & 1.96 &  26.3 &   5.68 & 12.25 &   1.84 & 0.67 \\  
 122 &   0.66 &   101.1 & 13:34:28.56 & 37:47:07.05 & -1.14 &  1.00 & 1.52 &  24.5 &   2.95 &  7.80 &   2.77 & 0.03 \\  
 130 &   0.63 &    72.0 & 13:34:44.09 & 37:44:34.90 & -0.01 & -1.52 & 1.52 &  26.9 &   6.85 & 10.25 &   0.92 & 0.62 \\  
 130 &   0.63 &    72.0 & 13:34:44.09 & 37:44:34.90 &  1.26 & -0.73 & 1.45 &  24.5 &   6.85 & 10.25 &   1.93 & 0.12 \\  
 149 &   0.50 &    80.1 & 13:34:01.19 & 37:53:49.14 & -0.48 &  1.35 & 1.43 &  22.8 &   5.68 &  7.12 &    --- &  --- \\  
 154 &   0.47 &    39.9 & 13:35:34.89 & 37:50:28.74 & -0.89 &  1.65 & 1.87 &  25.6 &   5.37 & 12.20 &   1.66 & 0.91 \\  
 162 &   0.43 &    30.6 & 13:34:33.62 & 38:05:40.27 & -0.61 &  0.04 & 0.61 &  25.1 &   7.71 & 10.96 &   1.78 & 0.77 \\  
 162 &   0.43 &    30.6 & 13:34:33.62 & 38:05:40.27 &  1.73 & -0.65 & 1.85 &  25.3 &   7.71 & 10.96 &   2.46 & 0.11 \\  
 170 &   0.37 &    78.8 & 13:34:17.03 & 37:59:49.48 &  1.25 &  1.06 & 1.64 &  22.9 &   1.02 &  6.43 &   1.12 & 0.91 \\  
 173 &   0.37 &    63.1 & 13:34:42.90 & 37:52:04.09 & -1.36 & -1.32 & 1.89 &  26.6 &   5.07 &  2.91 &   1.51 & 0.54 \\  
 192 &   0.27 &    57.6 & 13:34:21.56 & 37:49:59.82 & -0.83 &  1.51 & 1.72 &  24.7 &   0.46 &  5.63 &   1.77 & 0.57 \\  
 200 &   0.24 &    41.7 & 13:35:09.62 & 37:50:11.89 &  0.81 & -1.47 & 1.68 &  23.6 &   0.66 &  7.88 &    --- &  --- \\  
 202 &   0.23 &    27.2 & 13:35:20.44 & 37:55:29.25 &  0.83 & -1.74 & 1.92 &  25.9 &   5.37 &  8.60 &   2.10 & 0.35 \\  
 204 &   0.23 &    26.7 & 13:34:43.12 & 38:00:20.47 & -1.48 &  1.23 & 1.93 &  25.8 &   3.72 &  5.74 &   2.15 & 0.49 \\  
 210 &   0.19 &    26.0 & 13:34:33.22 & 37:52:22.40 & -0.13 &  0.99 & 1.00 &  20.6 &   3.75 &  2.48 &  14.05 & 0.03 \\  
\end{tabular} 
\end{center} 
\end{table*}

To avoid repetition below, when not clearly stated otherwise, `more
distant/closer' means more distant/closer to the {\em {\em Chandra}}
centroid and `brighter/fainter than' means `brighter/fainter than the
primary candidate listed in Table~\ref{tab:main}.  The sources are
labelled either `U', when the true identification is quite uncertain
and `P' when the primary identification is probably correct but {\sc
SExtractor} did pick up another, less likely, candidate within $2''$
of the \ch centroid.The primary candidate is chosen, unless stated
otherwise, on the basis of being closer to the {\em Chandra} centroid
and being brighter.

\noindent
{\bf 47 U} The alternate counterpart is $0.7''$ more distant
and 0.8mag fainter, although more stellar.\\
{\bf 56 U} The alternate counterpart is 0.7mag fainter and much more diffuse, 
although $0.3''$ closer.\\
{\bf 57 U}  The alternate counterpart is 1.7mag fainter  and $0.3''$
more distant. 
Neither counterpart is particularly close to
the {\em {\em Chandra}} centroid but the source is $10.7'$ from the {\em Chandra} axis.\\
{\bf 85 P} The alternate counterpart is 0.9mag fainter  and 
$1.6''$ more distant.\\
{\bf 95 P} The alternate candidate is 3.4mag fainter and much more distant
($1.6''$).\\
{\bf 102 U} In this case the identification is completely uncertain. 
The alternate candidate is 0.3mag brighter, but $0.25''$ more distant, 
and is less stellar.\\
{\bf 115  U} Both candidates are very faint (26mag) and so could be chance
coincidences. The alternate candidate is $1.3''$  more distant.\\
{\bf 122 P} The alternate candidate is 0.5mag fainter, more
diffuse, and $0.9''$ more distant.\\
{\bf 130 U} The identification may be with a cluster. The alternate candidate
is 0.2mag brigher, but $0.7''$ more distant.\\ 
{\bf 149 U} The alternate candidate is 0.9mag brighter but $0.8''$
more distant.\\
{\bf 154 P} The alternate candidate is 1.3mag fainter, and $1.4''$ 
more distant.\\
{\bf 162 U} Both candidates are at similar distances from the {\em Chandra}
centroid. The alternate candidate is 1.3mag fainter. However
the identification may be with a cluster as other objects are nearby.\\
{\bf 170 U} The alternate candidate is $1.6''$ more distant and 0.4mag
fainter.\\ 
{\bf 173 U} The alternate candidate is $1.5''$ more distant and 3.5mag
fainter.\\ 
{\bf 188 U} {\sc SExtractor} finds no candidate within $2''$. However
examination of the optical image shows a bright galaxy whose centre is
over 3'' to the NW. That is too far to be a reasonable
identification. However to the SE, and also to the NW of the galaxy,
there are bright components.  We cannot tell if these 2 components are
part of the same structure, eg a bright disc, seen edge on, or whether
they are separate objects, eg QSOs. The SE component lies very close
to the Chandra centroid and so its position is listed in
Table~\ref{tab:main}.\\
{\bf 192 U} The identification may be with a cluster. The nominally primary
candidate is $1.2''$ closer but 1.2mag fainter.\\
{\bf 200 U} This source is almost on axis so its position will be very 
accurate. Hence we list, as the primary identification, an object which
is 2mag fainter than, but $1.2''$ closer to the Chandra centroid than,
the alternative candidate.\\ 
{\bf 202 U} The alternate candidate is 4mag fainter and $0.7''$ 
more distant.\\
{\bf 204 U} Both candidates have similar magnitudes but the primary candidate
is $1.6''$ closer.\\
{\bf 210 U} There are 2 almost equally likely candidates, at similar 
distances. On the basis that it is a relatively bright radio source we list
the fainter, by 2.8mag, and more stellar, as the slightly more  
likely candidate.

\section{The Definition of $f_{x}/f_{opt}$}

The X-ray/optical ratio has been defined, historically, as the ratio
of fluxes in particular bands rather than at particular monochromatic
frequencies. In addition the $R$-band is now more widely used in deep
surveys than the $V$-band in which the ratio was originally
defined. It is therefore not always easy to compare ratios quoted by
different observers using different instrumentation.

The most widely used definition of $f_{x}/f_{opt}$ is that used by
Stocke \etal (1991), in conjunction with observations from the
Einstein Observatory, which is

\[ \log\frac{f_{x}}{f_{opt}} = \log f_{0.3-3.5} +\frac{V}{2.5} +5.37 \] 

\noindent
where $V$ is the Johnson magnitude, $f_{0.3-3.5}$ is the observed 
$0.3-3.5$ keV X-ray flux in units of \ecs and all logs are to base 10.

Here our fluxes are measured in the $0.5-7$ keV band and our optical
observations were made in the Cousins $R$ band.  To obtain an optical
flux we use the Cousins $R$-band (Vega) $F_{\lambda}$ normalisation
given by Fukugita \etal (1995) of $2.15 \times 10^{-8} {\rm W m}^{-2}
\mu^{-1}$ and multiply by the FWHM bandwidth, given by Fukugita
\etal (1995) as 0.1568$\mu$. We therefore conclude that

\[ \log\frac{f_{x}}{f_{opt}} = \log f_{0.5-7} +\frac{R}{2.5} - C \] 

\noindent
where C=9.53, $R$ is the Cousins $R$ magnitude, $f_{0.5-7}$ is the
observed $0.5-7$ keV X-ray flux in units of $10^{-15}$ \ecs.

Converting the Stocke \etal relationship to fluxes in the $0.5-7$ keV
band, in units of $10^{-15}$ \ecs we obtain

\[ \log\frac{f_{x}}{f_{opt}} = \log f_{0.5-7.0} +\frac{V}{2.5} - K \] 

\noindent
where K=9.82 for $\alpha=0.4$ and K=9.74 for $\alpha=0.7$.

Thus our relationship and that of Stocke \etal give the same value of
$f_{x}/f_{opt}$ for $V-R = 0.73$ (assuming $\alpha=0.4$) or 0.53
(assuming $\alpha=0.7$). These values of $V-R$ are not unreasonable
for extragalactic X-ray sources and so we conclude that we may use our
values of $f_{x}/f_{opt}$, in comparison with those of Stocke \etalc,
to make simple deductions about the nature of the X-ray sources.

We note that Giacconi \etal (2002), although measuring fluxes in the
same $0.5-7$ keV band as us, define $f_{x}/f_{opt}$ as $f_{0.5-10
{\rm keV}}/f_{R}$ and extrapolate their fluxes to the $0.5-10$ keV band
assuming an energy spectral index, $\alpha = 0.375$, ie a $0.5-7$ keV
flux of $0.76 \times 10^{-15}$ \ecs corresponds to a $0.5-10$ keV flux
of $1.0 \times 10^{-15}$ \ecs$\!$.  At the latter flux, fig.~16 of
Giacconi \etal gives $R\sim24.5$ for $f_{x}/f_{opt}=1$ whereas, at
$0.76 \times 10^{-15}$ \ecs ($0.5-7$ keV) our relationship gives
$R=24.1$.  This magnitude offset is not greatly changed by a steeper
choice of spectral index, such as may be more appropriate to our
slightly brighter sources, but may be due to a different definition of
$R$-band flux. The offset is not large, but may be borne in mind when
comparing our results (Fig.~\ref{fig:magflux}) with those of Giacconi
\etal  Above a flux level of $1 \times 10^{-15}$ \ecs (0.5-10 keV), our
$f_{x}/f_{opt}$ distribution is very similar to that of Giacconi \etal
and of the {\em ROSAT} distribution of Schmidt \etal (1998). We do not
sample the fainter flux levels where Giacconi \etal find many objects
of low $f_{x}/f_{opt}$.

\bsp

\label{lastpage}

\end{document}